\documentclass[10pt,acmsmall,screen]{acmart}
\usepackage{lang}

\usepackage{amsthm,amsmath,amssymb,latexsym}
\usepackage{hhline}    

\usepackage{adjustbox}
\usepackage{caption}
\usepackage{subcaption}
\usepackage{MnSymbol}
\usepackage{paralist}

\usepackage{booktabs}   
\usepackage{subcaption} 

\makeatletter\if@ACM@journal\makeatother
\setcopyright{rightsretained}
\acmPrice{}
\acmDOI{10.1145/3133892}
\acmYear{2017}
\copyrightyear{2017}
\acmJournal{PACMPL}
\acmVolume{1}
\acmNumber{OOPSLA}
\acmArticle{68}
\acmMonth{10}

\startPage{1}
\else\makeatother
\acmConference[PL'17]{ACM SIGPLAN Conference on Programming Languages}{January 01--03, 2017}{New York, NY, USA}
\acmYear{2017}
\acmISBN{978-x-xxxx-xxxx-x/YY/MM}
\acmDOI{10.1145/nnnnnnn.nnnnnnn}
\startPage{1}
\fi

\usepackage{lineno}
\usepackage{url}

\usepackage{tikz}  
\usetikzlibrary{
	shapes.geometric,
	arrows,
	shapes.symbols,
	shapes.misc,
	shapes.arrows,
	calc,
	chains,
	decorations.pathmorphing,
	decorations.pathreplacing,
	decorations.markings,
	matrix,positioning,
	shadows,
	positioning,
	trees
}
\tikzset{
	label/.style={
		font=\footnotesize\itshape
	},
	system/.style={
		font=\footnotesize,
		align=center,
		rounded corners=0.2em,
		draw,
		minimum height=2em,
		minimum width=4.6em,
		text height=1.5ex,
		text depth=0.01em,
    drop shadow,
    fill=white
	},
	monitor/.style={
			rectangle,
			rounded corners=0.2em,
			font=\footnotesize,
			minimum width=4.2em,
			minimum height=2em,
			text centered,
			draw,
			drop shadow,
			fill=white
	},
  process/.style={
			rectangle,
			rounded corners=0.2em,
			font=\footnotesize,
			text centered,
			draw,
			drop shadow,
			fill=white
	},
	smallsystem/.style={
		system,
		minimum height=2em,
		minimum width=3em,
		font=\footnotesize
	},
	squatarrow/.style={
		single arrow,
		draw,
		fill=black,
		minimum width=3em,
		minimum height=2.5em,
		inner sep=0ex,
		single arrow head extend=0.2em
	},
    tape/.style={
        font=\footnotesize\itshape,
        minimum size=1.4em,
        draw,
        text centered,
        text height=1.5ex,
        text depth=.25ex
    },
    tracer/.style={
        font=\footnotesize\sffamily,
        rectangle,
        rotate=90,
        minimum height=0.6em,
        minimum width=1.6em,
        text=white,
        fill=black,
        drop shadow
    },
    cnumb/.style={
        circle,
        align=center,
        minimum size=0.8em,
        inner sep=0.05em,
        outer sep=0em,
        font=\scriptsize\sffamily,
        draw,
        text=white,
        fill=black
    }
}

\newcommand{\invokedynamic}{\sv{invokedynamic}}

\input{mymacros.sty}

\begin{document}

\setlength{\pdfpageheight}{\paperheight}
\setlength{\pdfpagewidth}{\paperwidth}

\title{Heaps Don't Lie: Countering Unsoundness with Heap Snapshots}

\begin{abstract}
  Static analyses aspire to explore all possible executions in order to achieve soundness.
  Yet, in practice,  they fail to capture common dynamic
  behavior. Enhancing static analyses with dynamic information is a
  common pattern, with tools such as Tamiflex. Past approaches,
  however, miss significant portions of dynamic behavior, due to
  native code, unsupported features (e.g., invokedynamic or lambdas
  in Java), and more.  We present techniques that
  substantially counteract the unsoundness of a static analysis, with
  virtually no intrusion to the analysis logic. Our approach is
  reified in the HeapDL toolchain and consists in taking whole-heap
  snapshots during program execution, that are further enriched to
  capture significant aspects of dynamic behavior, regardless of the
  causes of such behavior. The snapshots are then used as extra inputs
  to the static analysis. The approach exhibits both portability and
  significantly increased coverage. Heap information under one set of
  dynamic inputs allows a static analysis to cover many more behaviors
  under other inputs. A HeapDL-enhanced static analysis of the DaCapo
  benchmarks computes $99.5\%$ (median) of the call-graph edges of
  unseen dynamic executions (vs. $76.9\%$ for the Tamiflex tool).

%
\end{abstract}




\author{Neville Grech}
\affiliation{
  \department{Dept. of Informatics}
  \institution{University of Athens}
  \streetaddress{Ilisia}
  \city{Athens}
  \postcode{15784}
  \country{Greece}
}
\affiliation{
  \institution{University of Malta}
  \country{Malta}
}
\email{me@nevillegrech.com}

\author{George Fourtounis}
\affiliation{
  \department{Dept. of Informatics}
  \institution{University of Athens}
  \streetaddress{Ilisia}
  \city{Athens}
  \postcode{15784}
  \country{Greece}
}
\email{gfour@di.uoa.gr}

\author{Adrian Francalanza}
\affiliation{
  \department{Dept. of Computer Science}
  \institution{University of Malta}
  \streetaddress{Msida}
  \country{Malta}
}
\email{adrian.francalanza@um.edu.mt}

\author{Yannis Smaragdakis}
\affiliation{
  \department{Dept. of Informatics}
  \institution{University of Athens}
  \streetaddress{Ilisia}
  \city{Athens}
  \postcode{15784}
  \country{Greece}
}
\email{yannis@smaragd.org}
\thanks{Authors' email: \textsf{me@nevillegrech.com}, \textsf{gfour@di.uoa.gr},
  \textsf{adrian.francalanza@um.edu.mt} and \textsf{yannis@smaragd.org}}

\maketitle

\section{Introduction}

Static analysis approaches typically attempt to be over-approximate
and cover all possible program behavior: when there are two possible
paths of execution, a static analysis explores both; when there are
many possible values for a variable, a static analysis examines all of
them, usually by employing an abstraction that groups together a
large number of concrete values.

Still, practical static analyses routinely suffer from
\emph{unsoundness}~\cite{soundiness15}, by failing to account for
standard dynamic behavior. The causes of this unsoundness are features
such as reflection, native code, dynamic loading, but also
cross-language development (e.g., hybrid Java-Javascript apps or languages
running on top of the JVM and integrating with the Java libraries) and the
engineering complexity of supporting a growing number or
more-and-more complex language features, such as Java's
\invokedynamic{} instruction. The typical modern Java application
uses complex frameworks that integrate external resources (e.g., XML
files) with inversion-of-control patterns that present static analysis
frameworks fail to account for.

An approach for coping with the ever-increasing dynamism of realistic
programs is to capture dynamic behavior and encode it as an input for
subsequent static analysis. For instance, \citet{ecoop/HirzelDH04,toplas/HirzelDDH07} attempt to counter
dynamic loading by observing its effects, recording the results, and
re-running the static analysis. The Tamiflex
tool~\cite{icse/BoddenSSOM11} records the result of reflective
operations and dynamic loading actions, produces a log as an input to
the static analysis, or even rewrites the program with these sources
of dynamic behavior replaced by the exact behavior observed during the
dynamic run.

Although these efforts have pushed the state of the art, they still
fall short of capturing many sources of unsoundness, such as program
semantics expressed in different languages (be it Javascript code for
UI elements, or C/C++ code in native libraries) or the lack of support
for cutting-edge language features (e.g., \invokedynamic{} and
lambdas). Unsound handling of such features translates into reduced
analysis \emph{coverage}: the static analysis misses many valid
program behaviors.

Our work proposes an approach that compensates for the coverage
shortcomings of static analysis by integrating dynamic information
produced from \emph{heap dumps}: snapshots of dynamic behavior that
record the shape of the heap, the stack shape (i.e., full stack
traces) when every object was created, and more. Heap dumps reflect a
substantial portion of the complex dynamic behavior of a program,
regardless of the cause of such behavior: instead of watching what
happens at specific \emph{actions} (e.g., reflection or dynamic loading
operations), a heap dump records the cumulative
semantic \emph{effects} of program execution in its native setting and
complex environment. At the same time, heap dumps do not miss the
ability to capture dynamic actions (e.g., a dynamic call-graph) since
each object (either natively or through \emph{heap enricher}
functionality that we introduce) records information describing the
dynamic context at the time of its allocation.

We implemented our approach in the HeapDL tool for Java programs, on
both the JVM and Android. HeapDL leverages different APIs to produce
standard HPROF heap dumps, and processes them to produce
representations of the heap and call graph that static analysis can
use. (HeapDL also produces a packaged version of both the statically
available and the dynamically loaded classes of the program.) We show
the benefits of HeapDL by importing its output in standard static
analyses (points-to and call-graph analysis). The result
demonstrates the benefits of our approach:

\begin{itemize}
\item Heap dumps produce significant increases in analysis coverage, compared
  to past techniques that enhance a static analysis with knowledge
  about dynamic actions (e.g., reflection and dynamic loading).
  A static analysis enriched by our HeapDL tool discovers 24\% more
  call-graph edges and 86\% more references between heap objects, compared to
  the same analysis enriched by the Tamiflex tool. The benefit is
  clear in direct comparisons of the predictive power of each analysis:
  given the same dynamic input, the HeapDL-enhanced analysis statically
  computes 99.5\% (median) call-graph edges of dynamic executions under
  \emph{different} inputs, vs. 76.9\% for Tamiflex.

\item HeapDL heap dumps are better suited for precise integration
  with static analysis. Our heap enricher allows the recording of
  context that is otherwise not readily available. Notably, we
  generate context-sensitive information for \emph{object-sensitive}
  analyses, of any context depth. This information can be directly
  integrated in a static analysis that uses the same context abstraction.

\item Heap dump technology is by nature more portable than dynamic
  agents that watch specific program actions. HeapDL supports both
  JVM and Android dynamic analysis, unlike past tools (e.g., Tamiflex)
  that are JVM-only. We argue that this is an inherent difference,
  rather than an outcome of current technology trends: it is more
  likely for a runtime environment to support snapshots of
  \emph{state} rather than arbitrary recording of program
  \emph{actions} during execution.
\end{itemize}

More generally, our approach follows a theme well-established in the
literature: the combination of static and dynamic analysis, so that
concrete information can take the place of static abstractions that
are hard or impossible to compute. In this general theme, there are
specific elements of our techniques that are unique, and are largely
responsible for the benefits we obtain. These elements include:
\emph{(a)} the use of heap snapshots with state-of-the-art technology;
\emph{(b)} the enhancing of such snapshots with extra context
information and with objects that would normally not be available;
\emph{(c)} the packaging of dynamic information for reuse by common
whole-program static analyses (such as points-to
analyses or call-graph construction). We next describe our approach
with an emphasis on these unique elements.



\section{Overview of the Approach}

We begin with an overview of the main elements of the HeapDL
approach: the overall workflow, current heap dump technology, and
output for integration with static analyses. The discussion in this
section is purposely simplified. In Section~\ref{sec:enriching} we
discuss how we enhance the basic scheme.

\subsection{Motivation and Main Idea}

The HeapDL approach consists of taking snapshots of a running
program's heap and using them to provide further input for a static
analysis. The intent is to uniformly capture the state-changing
effects of hard-to-analyze features. These features include native and
other heterogeneous code, cutting-edge language features, dynamic
loading, and more. A modern application crucially depends on such
features, yet static analysis frameworks (such as the Soot
infrastructure~\cite{vall99soot} or the Doop pointer analysis
framework~\cite{oopsla/BravenboerS09}) have incomplete support for
them. Examples in the Java world include:

\begin{itemize}
\item Virtually all modern Java programs have semantics that depend on
  native code. For instance, atomic operations are essential for
  high-performance shared-memory parallelism. Atomic reads and writes
  on the heap (e.g., to object fields or array entries) are
  implemented as native Java methods. If the static analysis does not
  model all of them, it will miss significant state updates. It is as
  essential for an analysis to model, e.g., native method
  \sv{sun.misc.Unsafe.compareAndSwapObject} as it is to support plain
  heap load and store instructions. Yet doing so is hard. Extra native
  operations get added in every release of the JDK and analysis
  authors typically do not keep up with them. These operations can be
  much more numerous than JVM instructions. On a quick count, there
  are over 6,000 native methods in OpenJDK 8u60 (vs. under 200
  instruction opcodes in the JVM instruction set).

\item Most enterprise or mobile Java programs heavily leverage complex
  frameworks, effectively becoming heterogeneous applications.  For
  example, an Android app is a complex composition of UI elements,
  whose specification is in XML, and Java code. Upon loading, the XML
  specification is used to instantiate many graphical components,
  which can also be referred to from plain Java code via dynamic
  lookups (using integer keys). Similar examples of framework usage
  abound---in Servlet coding, J2EE applications and more. Java
  Enterprise frameworks heavily employ XML specifications, with
  inversion-of-control patterns used to determine how plain Java code
  is invoked. Static analyses attempt to capture the semantics of such
  frameworks to the extent possible. E.g., the
  FlowDroid~\cite{Arzt14} add-on to the Soot framework implements
  basic processing of Android XML layout files. Yet such support
  is always vastly incomplete (as will also be apparent in our
  experimental evaluation) due to the complexity and ever-changing
  nature of modern frameworks.

\item Even with the limited size of the JVM instruction set, static
  analyses do not fully support it. Java 7 introduced a new bytecode
  opcode, \invokedynamic{}~\cite{Rose09}, together with an API (for
  ``method handles'') around it, that can offer the programmer the
  capability to completely customize dynamic program behavior. The
  \invokedynamic{} functionality is used to implement dynamic languages
  on the JVM and also a growing number of dynamic features of Java
  (e.g., lambdas~\cite{lambdas}, string concatenation~\cite{jep280},
  or generics specialization\cite{Goetz16}). To this date, support for
  \invokedynamic{} in static analysis frameworks has been, at best,
  incomplete.

\end{itemize}

\begin{figure}
  \centering
  \includegraphics[width=0.95\textwidth]{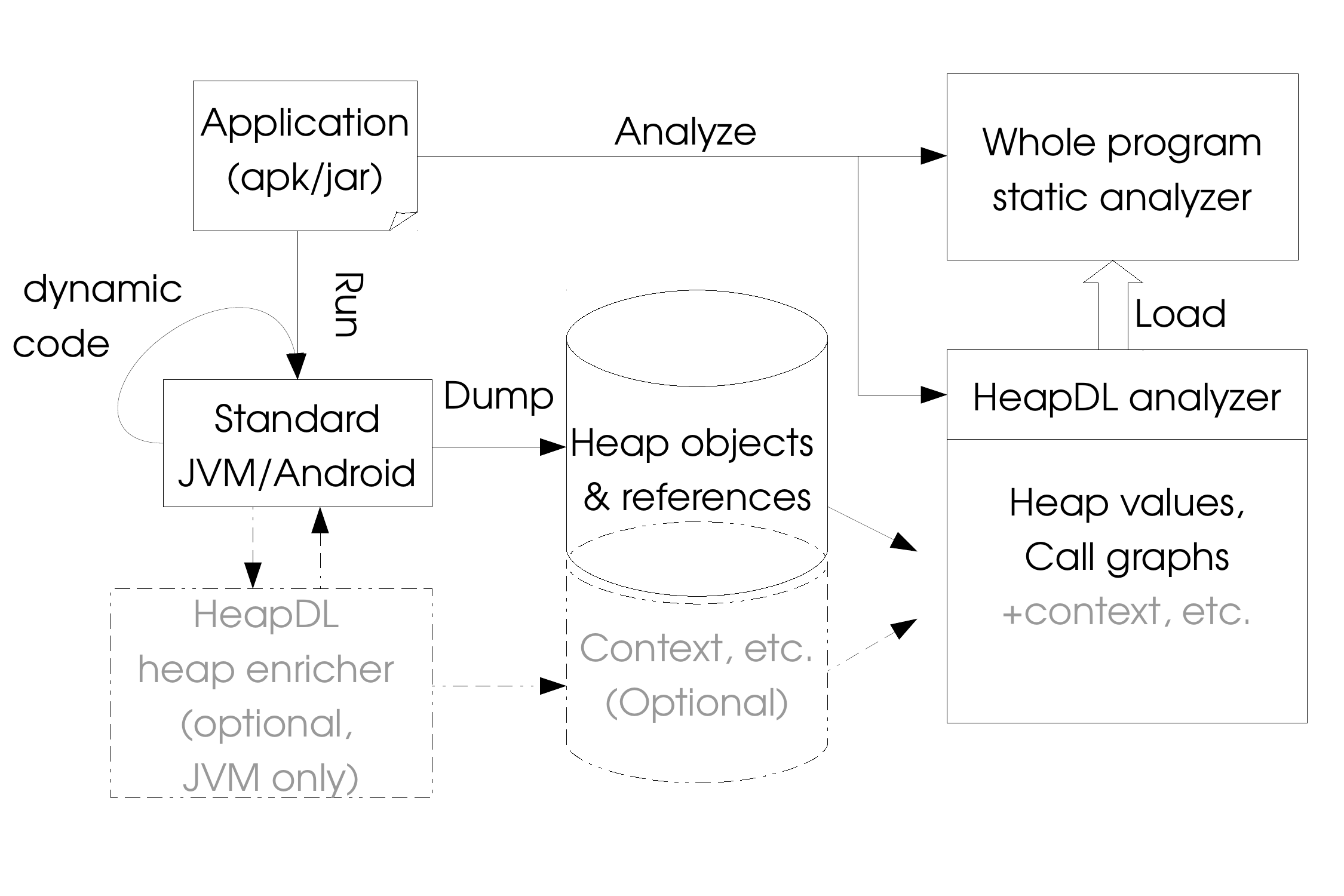}
  
%
%
%
%
%
%
\caption{Design of HeapDL}
\label{fig:heapdl-schematic}
\end{figure}

All the above instances result in \emph{unsoundness}; the static
analysis fails to capture actual dynamic behavior. This unsoundness
is quantified as reduced \emph{coverage} of program behavior.
HeapDL compensates by adding dynamic information to
static analysis.  Semantic effects, captured by the heap state and
dynamic call-graph of the application, are extracted from a heap dump
and used to supplement a static
analysis. Figure~\ref{fig:heapdl-schematic} shows the main
components, schematically. HeapDL relies on profiling capabilities
of the target runtime. Both major Java-based platforms, Android and
the JVM, provide multiple memory profiling and heap dumping solutions.
With an \emph{enriching agent} (Section~\ref{sec:enriching}) we can
make a heap dump encode even more information that is of direct
value to static analysis.

The dynamic information is output in a form suitable to import in a
static analysis. HeapDL explicitly targets whole-program analyses,
rather than local static analyses. It is, for instance, much better
suited for points-to analysis, heap shape analysis, or call-graph
construction, rather than for symbolic execution or model checking.
This is reflected both in the choice of technology for capturing
dynamic information (Section~\ref{sec:heapdump}) and in the packaging
of information for reuse (Section~\ref{sec:schema}).

To see how a heap snapshot can counter the effects of unsoundness in
static analysis, we can consider some concrete examples.


\paragraph{Example: external code effects.}

Consider an Android application, with several Java components linked
together by means of an XML specification. By taking a snapshot of the
running application, HeapDL can capture behaviors that are very hard to
follow via static analysis alone. For instance, the instantiation of
UI components and their inter-linking (e.g., a window object contains
a reference to three panels and a slider) will be hard to detect
statically, since it is implemented deep in the Android runtime, in
large part in native code. A heap snapshot can inform the static
analysis about the instances of these UI components and their
inter-connectivity. In this way, the analysis starts from a valid
initial setup and can cover substantially more code (e.g., by
statically analyzing possible called methods on these components).

\paragraph{Example: better reflection analysis.}

Even if we focus only on reflection analysis, heap snapshots can offer
advantages compared to merely recording dynamic reflective
actions. Consider a program that holds a large array of $k \approx 1000$ class
names, initialized so that no static analysis can know its values
(e.g., read from an XML file). These class names can represent
different to-be-loaded components (e.g., plug-ins of a large
application). The class names can be used to call methods via
reflection. In a single execution, a small number, e.g., 3, 
distinct class names are used. Current state-of-the-art tools
for handling reflection, such as Tamiflex, watch dynamic reflective
actions and hence record the calls to the 3 classes'
methods. Therefore, a static analysis enhanced with Tamiflex output
can also analyze the 3 reflective calls.  HeapDL takes a heap
snapshot, so it can capture all $k$ members of the array. A
static analysis enhanced with HeapDL output and with minimal
reflection logic will analyze all possible calls to all $k$ classes'
methods.

\paragraph{Example: handling extra language features.}

Consider a static analysis that does not handle the \invokedynamic{}
instruction or its associated method-handles API. Heap snapshots
can alleviate the effects of such unsoundness in two ways. First,
a heap snapshot also includes snapshots of dynamic call graphs,
and can, therefore, capture the target of an \invokedynamic{} call.
Second, the heap effects of the method called via \invokedynamic{}
are captured in the snapshot. In this way, a static analysis enhanced
with HeapDL input can attain significantly higher coverage of
program behaviors that employ \invokedynamic{} calls.

Generally, heap snapshots can capture complex dynamic behavior
that is otherwise invisible to a static analysis and augment the
static analysis with such information.

\subsection{Background: Heap Dumps, Allocation Tracking}
\label{sec:heapdump}

HeapDL implements a heap dump analyzer that accepts standard
HPROF~\cite{hprof17} heap dumps. Java or Android applications are
dynamically executed by running on the unmodified, standard Java
Virtual Machine or Android runtime.

A heap dump is primarily a complete encoding of a program's heap as a
graph, i.e., a snapshot of all interconnections between heap objects.
Heap dumps can contain anything that is loaded or computed by the
application or VM, i.e., not just any normal heap objects constructed
by the application but also primitives, class objects, and strings.
This view, however, is too poor to capture the wealth of information
available through heap-dump APIs. In our setting, when we refer to a
heap dump, we mean a heap dump with \emph{allocation tracking}: each
heap object records a full stack trace of the run-time context at the
exact allocation instruction. Allocation tracking has a run-time cost
but is a portable facility, uniformly available in modern heap dump APIs.  By
leveraging allocation tracking, a typical heap snapshot also
integrates \emph{many thousands of stack snapshots, at earlier points
  of the execution} (i.e., whenever a heap object was
allocated). These stack snapshots are significantly condensed,
containing merely call-graph edges (i.e., which instruction called
which method) rather than full stack contents. (In
Section~\ref{sec:enriching} we see how we force the collection of even
more stack information, via our enriching agent.)  This is, however,
highly valuable information for enhancing the coverage of a static
analysis. Compensating for the unsoundness of reflection, dynamic
loading, \invokedynamic{}, inversion-of-control patterns, etc. is
majorly facilitated by these dynamic call-graph snapshots.

\subsection{Output Schema}
\label{sec:schema}

HeapDL accepts as input both the program code and a heap dump. It
then distills the heap dump into input tables for a static analysis, by
mapping objects and call-graph elements to abstractions. These
abstractions are derived by consulting the program code. HeapDL then
outputs the tables in standard text form, as comma-separated value
files, with appropriately externalized identifiers.

In Figure \ref{fig:domain-heap-insens}, we present a schema of the
domain of tables created by HeapDL for consumption by a
context-insensitive static analyzer. The heap relations generated by
HeapDL bridge the gap between the domain of an application's state and
the domain of static analysis. The relation
\predname{ObjectFieldValue} captures what values an object's fields
can point to, and similar information is kept for static fields of a
class (\predname{StaticFieldValue}) and arrays
(\predname{ArrayContentsValue}). \predname{CallGraphEdge} captures the
dynamic call-graph: every pair of successive stack trace elements
forms an edge. That is, the call-graph is the union of all call-graph
edges taken from the (large number of) stack snapshots collected due
to allocation tracking. Notably, we have found that call graphs
constructed in this way are comparable in size and information content
with ones created using explicit instrumentation of calls. (This is
perhaps not too surprising: object allocation is frequent and virtually
all meaningful call chains will reach code that causes at least one
allocation, possibly of a temporary object, resulting in the call
chain's capture.) Instrumentation, through Java or native agents, is
a less portable technique than heap snapshots, however---e.g., there
is no Android API for user-defined agents; bytecode rewriting can be
used but fails for native code or system classes.


All of the above mappings employ different kinds of abstraction:
objects are mapped to abstract objects, array contents are merged, the
union of field-points-to sets (per abstract object) or call-graph-edge sets
(per invocation site) is taken. Our heap object abstractions, $O$,
match those typically used by whole-program static analysis
frameworks, i.e., usually represent allocations sites:

\begin{javacode}
..
String[] a = new String[4]; // allocation site
Object o = new Object(); // allocation site
..
\end{javacode}

On the other hand, when statically modeling string constants and class
objects, the \emph{identity} of these is used as the object
abstraction, instead of their allocation site. For instance, the identity of
classes is the fully qualified name (and the classloader if the
static analyzer can distincguish classes with the same name loaded by different
classloaders).
The identity of strings can also be their content if the static analyzer is
tracking strings for the purpose of static reflection analysis.

\begin{figure}[tb!p]
\hspace{-2.6mm}
\begin{tabular}{lll}
\small $O$ is a set of object abstractions (e.g., allocation sites) & &
\small $F$ is a set of fields \\
\small $T$ is a set of class types & &
\small $I$ is a set of instructions \\
\small $M$ is a set of methods \\
\noalign{\vskip 1mm}
\pred{ObjectFieldValue}{obj : O, field : F, value : O} \\  
\pred{StaticFieldValue}{class : T, field : F, value : O} \\
\pred{ArrayContentsValue}{obj : O, value : O} \\ 
\pred{CallGraphEdge}{invocation: I, method : M} \\
\pred{Reachable}{method : M} \\
\end{tabular}
\caption[]{Our domain, for context-insensitive heap relations extracted by HeapDL}
\label{fig:domain-heap-insens}
\end{figure}

To generate dynamically inferred heap relations, HeapDL must first find
the right object abstractions from the heap dump. This is often a best-effort
match. HeapDL walks the allocation traces and uses heuristics to find the most
probable frame where the real allocation site is found as a first approximation. Given
this approximation, it tries to then match by type, line number, and other
information. Some of this information is only present in debug information of the
bytecode. Since the line number is not always guaranteed to be present in the
application under analysis, matching is sometimes done just by method descriptor
and type. In cases were the actual code is not statically available, a dummy
abstract object allocation site containing the right type information is
generated. This typically happens due to either native code, foreign code,
or cutting-edge language features such as lambda meta-factories that generate
transient classes and are incompletely modeled.


With the above schema, the information that HeapDL provides to static
analysis is compact and in line with current static (whole-program)
points-to analyses or call-graph construction.  A static analysis
typically only needs to import the HeapDL information and consider it
as ground facts, before it starts its own further propagation of
values. In general, the integration of HeapDL into an analysis
toolchain is similar to that of past tools, such as Tamiflex: dynamic
execution yields call-graph edges and object references, in an
externalized format (comma-separated value files). An
analysis-specific import method subsequently performs a
straightforward mapping from the externalized information to the
structures that the analysis uses to represent its own inferences.

Since static analysis is fundamentally over-approximate, small amounts of
provided information (e.g., a few hundred extra call-graph edges or
values in object fields that were previously undetected) are often
responsible for making the static analysis compute a much larger number of
inferences.



\section{Enriching Heaps and Context Sensitivity}\label{sec:enriching}
Enriching heaps is a process of strategically making small additions
to the state of the application so that a heap dump maintains more information
that we would like to produce as input to a static analysis. This technique
leverages the state-preserving abilities of the profiling toolchain.
It also preserves the actual linking between
objects and references of the original state of the application with that of the
additional information.

There are three main ways that the HeapDL context
enricher injects additional information into the state of the application:
\begin{itemize}
  \item Adding new references within agent code, e.g., during class loading.
  \item Injecting code into the application to add new references within the application.
  \item Injecting code into the application to create new objects at strategic
    program points.
\end{itemize}

These additions are typically made using \emph{instrumentation
  agents}, in Java or in native code, through standard APIs of the
JVM. HeapDL currently only supports heap enriching on the JVM
platform, since Android does not have a standard API for agents.


HeapDL performs heap enrichment for several different purposes, detailed next.

\subsection{Capturing Dynamically Loaded Code}
\label{sec:dynamic-code}

HeapDL captures all dynamically-loaded bytecode and packages it for
use by a static analysis. This is beneficial, since dynamically-loaded
classes (including temporary dynamically-generated code---e.g., for
\invokedynamic{} and other method-handle API support) would not
otherwise be available for static analysis. This general pattern has
also been present in past work. For instance,
Tamiflex~\cite{icse/BoddenSSOM11} creates an archive file with loaded classes by
instrumenting class-load events via an agent.

The complication, however, is that a loaded class is not uniquely
identified by its name (or its bytecode, as provided to the
loader). In a running JVM, a class's identity is represented by a
combination of its static identity (i.e., its name, which is an
artificial id for internally-generated, anonymous classes) and its
class loader object (an instance of type
\sv{ClassLoader})~\cite[\S{5.3}]{jvms}. The class loader can
arbitrarily transform the loaded bytecode. Therefore, an approach that
records bytecode by capturing the inputs of class-load events (before
actual loading has taken place) is incomplete: the
uniquely-identifying version of a class is only available after
loading is complete.

Enriched heap dumps can solve this problem and capture loaded classes
together with their instances on the heap. In a plain, un-enriched,
heap dump, objects do not refer to their classes' bytecode, as this is
compiled away by the VM. Instead, we can instrument the code (at the
point of loading classes) to perform a simple addition to the state so
that dynamically loaded bytecode is
captured. Figure~\ref{fig:enrich-bytecode-code} shows the
skeleton---we omit features such as error handling, logging,
performance optimizations, etc. for clarity.


\begin{figure}
\begin{javacode}
class ClassData {
  String name;
  ClassLoader loader;
  byte[] bytecode;
  ...
}

static List<ClassData> classes = new ArrayList<>();

public byte[] transform(ClassLoader loader, String name,
                        Class<?> clazz, .., byte[] bytecode) {
    ...
    classes.add(new ClassData(loader, name, bytecode));
    ...
}
\end{javacode}
\caption{Heap enricher example: Enriching heap with bytecode of loaded classes (code).}
\label{fig:enrich-bytecode-code}
\end{figure}

\begin{figure}
    \includegraphics[width=0.7\textwidth]{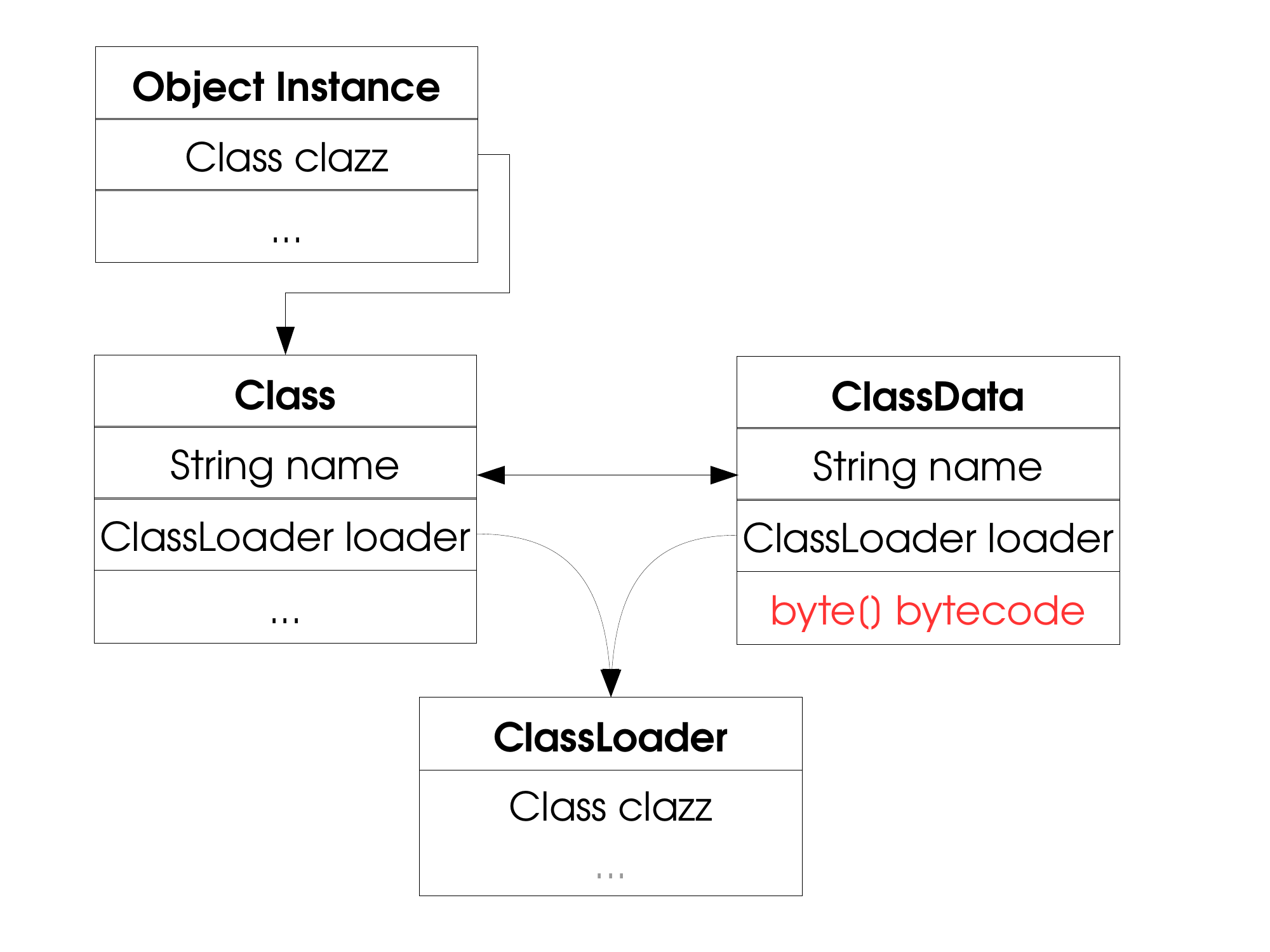}
\caption{Heap enricher example: Enriching heap with bytecode of loaded classes.}
\label{fig:enrich-bytecode-rel}
\end{figure}

We can see in Figure \ref{fig:enrich-bytecode-code} that capturing the
loaded bytecode can be achieved by storing, on line $13$, a reference
to the \sv{ClassLoader}, the fully qualified name of the class, and the
bytecode used to create it. With this technique there is no
need for extra logic to package the classes. We use this combination
of objects as keys and the structure of the heap dump contains links
from class object references to their bytecode via this key. Additionally,
the heap dump contains links from each instance object to its class object,
and with the unique name-loader combination we have the set of
associations shown in~Figure \ref{fig:enrich-bytecode-rel}: from
every object, we can get its (dynamic) class and bytecode.

\subsection{Context Sensitivity}

Whole-program static analyses often employ \emph{context sensitivity} to increase
precision.  Context sensitivity consists of qualifying all analysis
inferences with special ``context'' entities, so that different
dynamic executions are distinguished. The two main kinds of context
sensitivity are \emph{call-site sensitivity}~\cite{Sharir:Interprocedural}
and \emph{object sensitivity}~\cite{1044835}, with several alternatives and mixes
proposed (e.g., \emph{type sensitivity}~\cite{popl/SmaragdakisBL11}
and \emph{hybrid object-call-site sensitivity}~\cite{pldi/KastrinisS13}).

It is, therefore, desirable to provide context-sensitivity as an option
for HeapDL heap snapshots. Since HeapDL has full access to dynamic information,
it makes sense to preserve at least as much precision as the static analysis
seeks to achieve. This requires a) creating a general infrastructure to
instantiate and manipulate arbitrary context; b) capturing context not
usually present in heap dumps.

Heap dumps with allocation tracking require no extra effort to support
call-site sensitive contexts. In call-site sensitivity, context
consists of a tuple of call sites (i.e., invocation instructions) that
identify ``callers''. For a call-graph edge, the context of the called
method is its caller, the caller's caller, and so on, up to a maximum
context depth. Similarly, an allocated object's context is the caller
of the method that allocated it, the caller's caller, etc. This
information is naturally present in dynamic stack traces, which yield
information for relation \predname{CallGraphEdge} and for every
allocated object on the dynamic heap (via allocation tracking).

In contrast, object sensitivity is not possible to implement
from stack traces alone---our enriching agent has to maintain extra
information. Object-sensitive context is a tuple of \emph{abstract objects},
representing the receiver object of different calls. For a call-graph edge,
the context of the called method is its receiver (abstract) object, \emph{rec}; the
receiver, \emph{rec2}, of the method call that allocated \emph{rec}; the receiver
of the method call that allocated \emph{rec2}; and so on. Similarly, an
allocated object's context is the receiver of the method call that allocated it,
the receiver of the call that allocated the former receiver, etc.


Object sensitivity is both valuable in practical analyses and an
excellent example of our
heap enriching mechanisms for context sensitivity. We describe its
support next, on the two key parts of heap dump information: context for
heap objects and context for methods in a dynamic call-graph.
  
\subsubsection{Storing Heap Contexts on Object Creation}

In order to support object sensitivity, HeapDL 
maintains extra context information per allocated object. This is done
via a class \sv{ObjAndCtx} that associates each dynamic object with
its allocation context. HeapDL instruments the application
code to allocate instances of \sv{ObjAndCtx} every time a regular
object would be allocated. The HeapDL heap enricher is implemented as
a Java agent that performs load-time structured bytecode
transformations. This is by no means the only way to implement such a
strategy. Other ways include native agents, aspect oriented
programming with bytecode weaving, and more.

\begin{figure}
\includegraphics[width=0.95\textwidth]{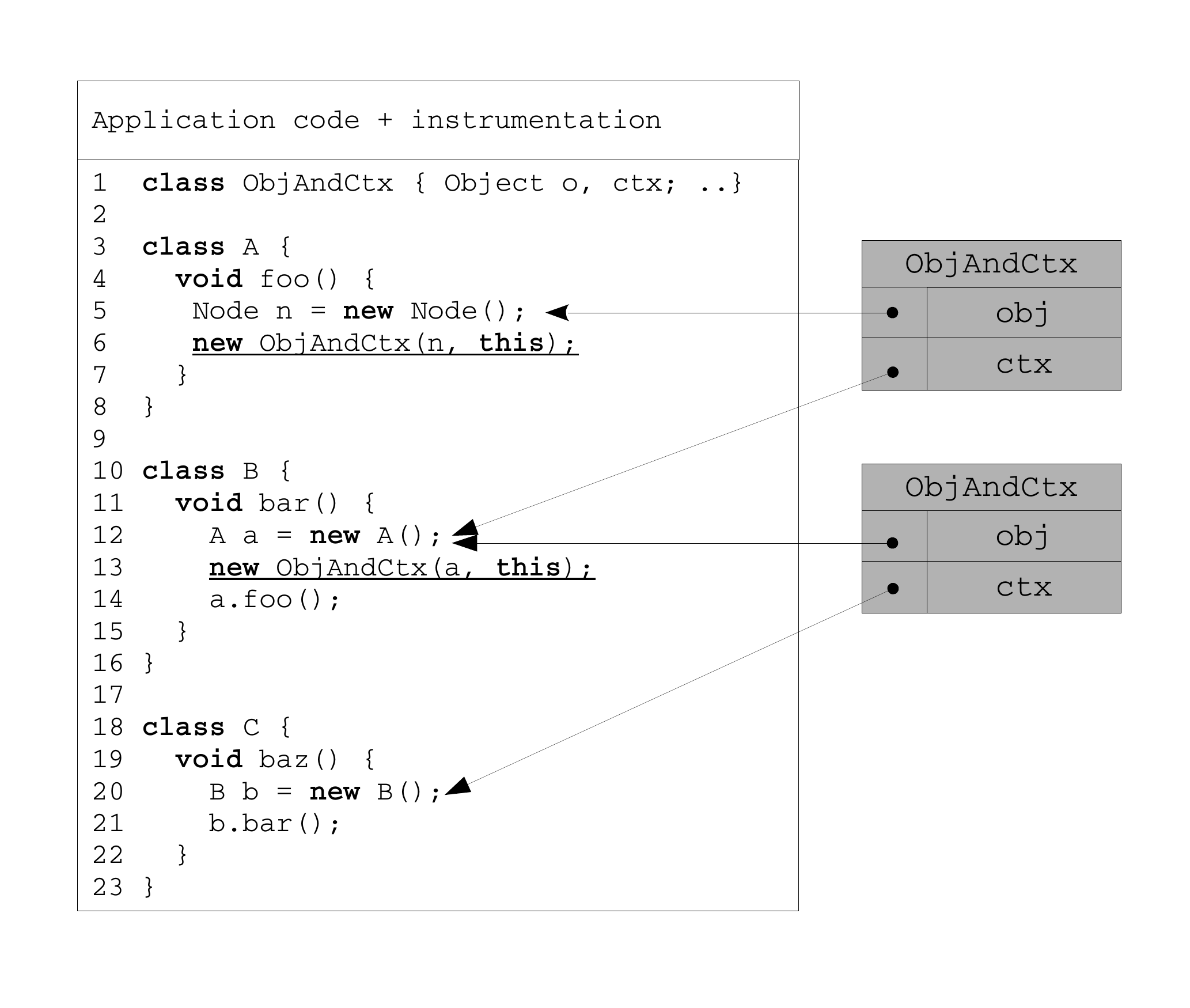}
\caption{Enriching the state with object sensitive heap contexts}
\label{fig:enrich-ctx-code}
\end{figure}

Figure \ref{fig:enrich-ctx-code} shows a target program with additional
instrumentation for object sensitivity. The tuple structure that
represents the heap context of an object appears at Line 1. This is injected
into the classpath of the application and instances of this structure are created on lines
6 and 13. At any interesting program points where an object is constructed, the
instrumentation keeps track of the receiver of the current method.

Note that, even though one receiver object is kept, object-sensitive context
of any depth can be computed: the receiver object contains a reference to
the receiver object of its own allocation method, etc.

The example shows objects constructed within instance
methods. Objects created inside static methods are handled
differently, since these have no receiver object, so the receiver of
the caller is used for the purposes of heap contexts. Every currently
active stack frame also keeps track of its receiver object, not shown
in the figure.

HeapDL keeps track of context information only inside application
code, to focus on the cases that require maximum precision, and to
avoid errors with the instrumentation of libraries. It also economizes
by not tracking the context of commonplace objects (which is typically
not statically modeled), such as primitive arrays, strings and string
buffers.


\subsubsection{Storing Calling Contexts for Context-Sensitive Call Graphs}

The second piece of information output by HeapDL that needs to be context-qualified
is call-graph edges.
%
%
Figure \ref{fig:enrich-edge-ctx-code}, shows an additional simple data
structure, \sv{EdgeCtx}, inserted into the application's class path and storing
the calling context of a call-graph edge. We can see that each
\sv{EdgeCtx} object contains a reference to the caller context and the callee
context. The \sv{storeCallerCtx} method is used by the instrumentation to keep
track of the receiver of the caller, which is then used during the creation
of the object by \sv{getCallerCtx}.

\begin{figure}
\begin{javacode}
class EdgeCtx {
  Object callerCtx, calleeCtx;
  
  void EdgeCtx(Object calleeCtx) {
    this.callerCtx = getCallerCtx();
    this.calleeCtx = calleeCtx;
  }

  static void storeCallerCtx(Object o) { ... }
  static Object getCallerCtx() { ... }
}
\end{javacode}
\caption{Edge Context}
\label{fig:enrich-edge-ctx-code}
\end{figure}

\begin{figure}
\includegraphics[width=0.95\textwidth]{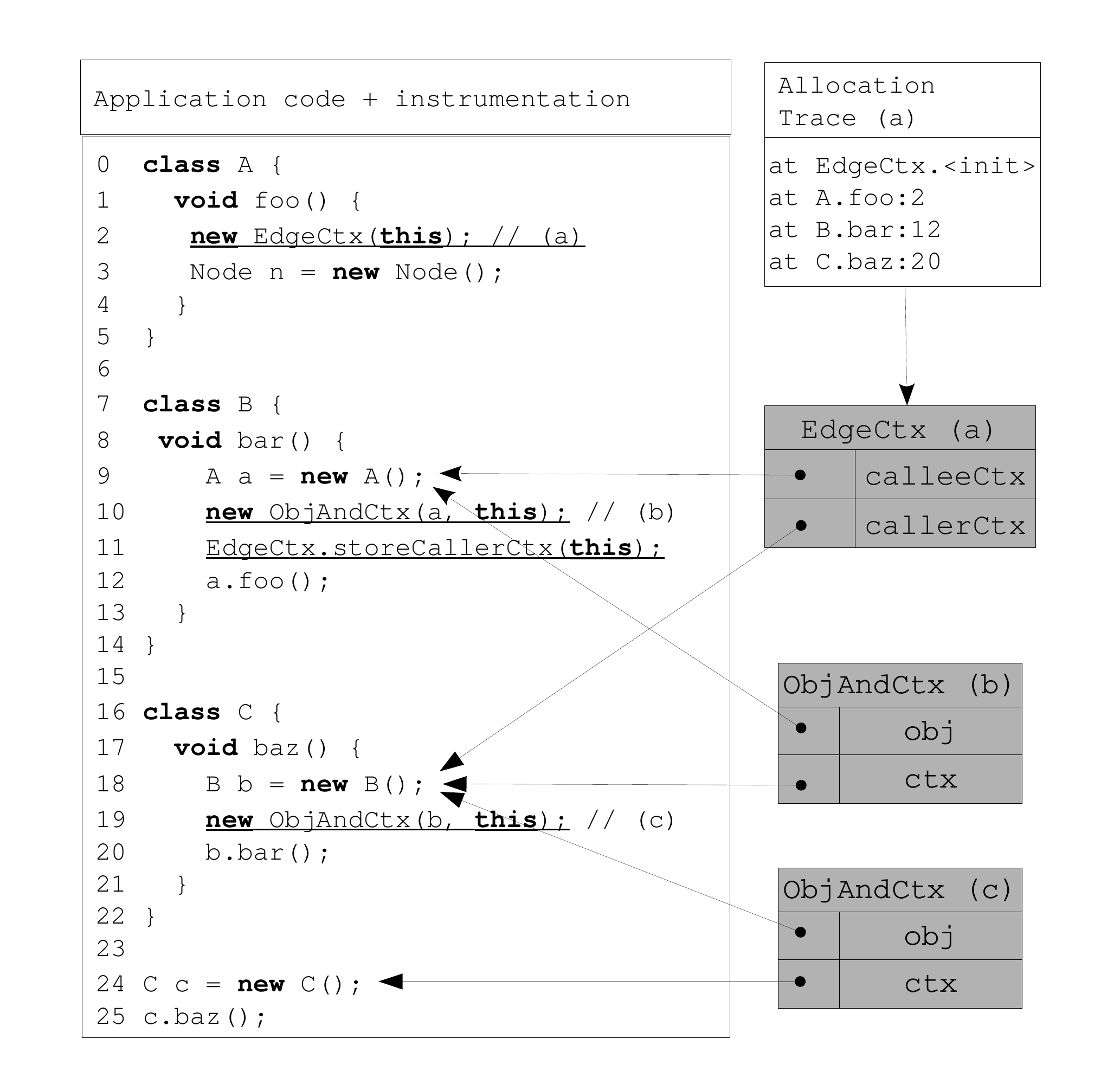}
\caption{Original application code, instrumentation by context heap enricher
  (underlined), and the depiction of the interaction between instrumentation heap
  objects (gray), allocation traces, and original heap objects.}
\label{fig:enrich-edge-ctx-diag}
\end{figure}

The enriching agent adds code to allocate a new \sv{EdgeCtx} at every
method call. When a new \sv{EdgeCtx} object is instantiated, a stack
trace is created, as illustrated in Figure
\ref{fig:enrich-edge-ctx-diag}.  In the stack trace, HeapDL can
extract the call-graph edge's source and target from the 2nd and 3rd
elements.
In the case of object sensitivity, as in our earlier discussion, the
explicit context pointer is to a single object, however one can
extract context of any depth by following the pointers to the
(context) objects and retrieving their own allocation context.


An interesting observation is that \sv{EdgeCtx} objects are by themselves
a representation of dynamic call-graph edges, and, indeed, the most precise
one. Without \sv{EdgeCtx} objects, a heap dump has no representation of
a specific dynamic call-graph edge, only of its mapping to source and target pairs,
as found in a stack trace (kept via allocation tracking). Unlike
the earlier \sv{ObjAndCtx} structures, which give context to dynamic objects (which
exist in the heap dump), \sv{EdgeCtx} instances cannot be uniquely mapped to
other heap dump entities.




\subsubsection{Producing Context Sensitive Information for Consumption by Static
  Analyzers}

Figure \ref{fig:domain-heap-sens} contains a refined version of the
domain and output relations extracted by HeapDL in Figure
\ref{fig:domain-heap-insens}. In order to show how our output relations are
built we need to further distinguish between abstract objects, $O$, i.e., the
objects used in the domain of static program analyzers, and concrete objects,
$L$. These concrete objects are ones that are found inside the heap dump,
including allocation-trace objects. Context-sensitive static analyses have
parametric order in their contexts, and the calling and heap context orders can
be different. (An example of this is a 2-object sensitive analysis with a 1-object
sensitive heap context.) In our domain, $n$ and $m$ are the orders of the calling
and heap context respectively.

In heap dumps analyzed by HeapDL, concrete contexts $L_C$ are a subset of
concrete objects $L$. Note how, in earlier examples (Figures~\ref{fig:enrich-bytecode-code} and
\ref{fig:enrich-edge-ctx-code}), all kinds of dynamic context shown were of type \sv{Object}.
This is done on purpose: dynamic information maintains full detail until the moment
it is packaged into output relations. Concrete contexts are then abstracted to abstract
context components, $C$. The process differs, depending on the kind of context
sensitivity that is employed. This process is defined using an abstraction
function $\alpha$. For example, in the case of object sensitivity, the
abstraction function used is the same as the abstraction function used to map any
concrete object in $L$ to an abstract object in $O$. In a type-sensitive analysis,
the abstraction function would yield the class in whose code the concrete object
got allocated.

Finally, in order to construct higher-order contexts, a function that maps
concrete objects to their contexts $\beta$ is applied recursively $n$ or $m$
times to get the required number of concrete components for a calling or heap
context respectively. In HeapDL, for an object- or type-sensitive analysis this
mapping is built from the information references found inside \sv{ObjAndCtx}
and \sv{EdgeCtx} objects.

\begin{figure}[tb!p]
\hspace{-2.5mm}
\begin{tabular}{ll}
  \small $L$ & Concrete objects \\
  \small $L_C : L$ & Concrete context components\\
  \small $C$ & Abstract context components \\
  \small $n : \mathbb{Z}^+$ & order of calling context sensitivity \\
  \small $m : \mathbb{Z}^+$ & order of heap context sensitivity \\
  \small $C_C : \bigtimes^n C$ & static calling contexts \\
  \small $C_O : \bigtimes^m C$ & static heap contexts  \\
  \cline{1-2}
  \small $\alpha : L_C \rightarrow C$ & context abstraction function \\
  \small $\beta : L \rightarrow L_C$ & concrete context component of concrete
  object \\
  \cline{1-2}
\end{tabular}
\begin{tabular}{ll}
  \noalign{\vskip 1mm}
  \pred{ObjectFieldValue}{ctx : $C_c$, obj : O, field : F,  hctx : $C_O$, value
    : O}\\
  \pred{StaticFieldValue}{class : T, field : F, hctx : $C_O$, value : O} \\
  \pred{ArrayContentsValue}{hctx$_{obj}$ : $C_O$, obj :  O, hctx$_{val}$ :
    $C_O$, value : O}\\
  \pred{CallGraphEdge}{callerCtx: $C_C$, invocation: I, calleeCtx : $C_c$,
    method : M} \\
  \pred{Reachable}{ctx : $C_c$,  method : M} \\
\end{tabular}
\caption[]{Our domain, for context-sensitive heap relations extracted by
  HeapDL}
\label{fig:domain-heap-sens}
\end{figure}

\subsection{Liveness}
The instrumentations and additional references, particularly those that capture
context information, tend to force many more objects to remain live (i.e.,
reachable from GC roots). Although this negatively affects the performance of
the application, it helps to increase the amount of information that can be
extracted for the heap. Therefore, even though the context-sensitive heap
enricher is not suitable to be used on live mission critical systems, it can be
used during pre-deployment analysis of an application with great benefits, even
for a context-insensitive analysis. Although during our experimental evaluation
we have not run into memory issues directly due to the heap-context agent, to minimize
the memory impact one can use the number of allocations referenced per
instrumentation point and randomly discard forced-live objects based on a
probability computed by a logarithmic function on the number of allocations.

\section{Discussion}

Before we evaluate experimentally the impact of the HeapDL approach,
it is useful to consider conceptually its properties, as contrasted with
Tamiflex---a state-of-the-art tool for handling dynamic language features.

The Tamiflex tool~\cite{icse/BoddenSSOM11} observes different run-time
events that pertain to dynamic language features, notably reflection
and dynamic loading. The outcome of such events is recorded, so that
static analysis can take it into account. For instance, a reflective
call can be treated as a regular call with a known target, corresponding
to the method observed in the instrumented run.

HeapDL uses heap snapshots that serve a dual purpose: they record both
dynamic \emph{events} as in the Tamiflex approach (through stack
traces kept via allocation tracking), and dynamic \emph{state}. When
this state is used as input to an over-approximating static analysis,
the results can model a lot more dynamic behaviors than those actually
observed during the profiled program run. This is a fundamental
feature of the approach: it captures not-seen behaviors and, as
expected in an over-approximating static analysis, some of them may be
spurious.

For a toy example, consider a method that performs a virtual call
over a value read from the heap:

\begin{javacode}
void m(X x) { 
  C c = x.f; 
  c.foo(); 
}
\end{javacode}

An actual program run may observe a call to method \sv{m} while
\sv{x.f} holds an object of type \sv{C1} (a subtype of \sv{C}).
An approach, such as that of Tamiflex, that watches dynamic
events will record the call-graph edge from \sv{m} to
\sv{C1.foo()}. However, the field \sv{x.f} may hold several
different values during the program's execution. Furthermore,
the over-approximating nature of static analysis may infer
that \sv{x.f} points to several different values during program
execution---e.g., even if each concrete object only holds a unique
value in its \sv{f} field, grouping concrete objects into abstract
ones can yield multiple values for \sv{x.f}.

Consider a case where, later in the program's execution, \sv{x.f}
acquires a second value: an instance of a class \sv{C2}, also a
subtype of \sv{C}. If a HeapDL snapshot captures that value, the
static analysis will consider it and will yield a call-graph edge from
\sv{m} to \sv{C2.foo()}. This does not correspond to any observed
program execution and may even be spurious: there is no guarantee that
method \sv{foo} is ever invoked on that value of \sv{x.f}.


Therefore, HeapDL can increase the analysis \emph{reach}: it models
more program behaviors than approaches that only observe dynamic
events. At the same time, it is interesting to evaluate whether the
increase in reach is reasonable (and not the result of vast
imprecision) as well as whether it corresponds to an increase in
\emph{coverage} of actual program behaviors.

\section{Experimental Evaluation}
In this section we present the results of an experimental evaluation of
HeapDL. This evaluation intends to answer the following research questions:

\begin{description}
\item [RQ.A] Is HeapDL effective?
  \begin{inparaenum}
  \item Does it expose new information that is not currently picked up
    through static analysis?
  \item What impact does this have on the results of the analysis?
  \item Does this gain also occur when explicit support for reflection is
    switched on in pointer analysis?
  \item Furthermore, does HeapDL find additional information that a
    state-of-the-art runtime analysis system like Tamiflex does not?
  \end{inparaenum}
\item [RQ.B] Is HeapDL efficient?
  \begin{inparaenum}
  \item What is the additional dynamic analysis burden (i.e., at program run time)?
  \item How much does the additional information add to the static analysis time?
  \end{inparaenum}
\item [RQ.C] Does HeapDL increase coverage of the analysis, compared to a
  state-of-the-art runtime analysis tool?
\end{description}

\newcommand{\rqstyle}[1]{\textit{#1}}

The size of the call graph, measured by the number of call graph edges, is used
as metric in \rqstyle{RQ.A} and \rqstyle{RQ.B}. In addition, we use the heap size as a metric in
\rqstyle{RQ.A}. The heap size is the cumulative size of relations, \predname{ArrayValue},
\predname{InstanceFieldValue}, and \predname{StaticFieldValue}, as described in Figure
\ref{fig:domain-heap-sens}.

The HeapDL analyzer is implemented\footnote{Available online at
  \url{https://github.com/plast-lab/HeapDL} and \url{http://heapdl.nevillegrech.com}}
as a plain Java application that produces tables in comma-separated-values format.
We used the Doop framework~\cite{oopsla/BravenboerS09} as a static analysis that
accepts HeapDL input.
HeapDL has integrations with two different implementations of Doop, that use
the Souffl{\'e}~\cite{Jordan16} and LogicBlox~\cite{Aref15} Datalog dialects, respectively. The integration
with Doop so it can import HeapDL information is minor, consisting of merely
importing data and considering them analysis facts. Doop has full support for
complex Java language features, such as class initialization, exceptions, reflection,
etc. In addition, Doop has recently acquired state-of-the-art support for
Android applications~\cite{Grech2017infoflow}.
It specifically models the Android lifecycle, callbacks, GUI
components, etc. Hence, enhancing Doop's coverage is not trivial.
To parse heap snapshots, HeapDL uses a modified version of
JHat~\cite{jhat}, the reference Java heap analysis tool supplied as part of
OpenJDK. The modifications consist of error recovery and the addition of class
pool information. For bytecode engineering, HeapDL's context heap enricher uses
the ASM framework~\cite{Bruneton02}---popular for tools that
manipulate or analyze Java bytecode.

Our runtimes are established on an idle machine with an Intel Xeon
E5-2687W v4 3.00GHz with up to 512 GB of RAM. For static analysis with Doop, we
used the PA-datalog engine, a publicly available, stripped-down
version of the commercial LogicBlox Datalog engine. We proceed in the
next sections with the experiments using popular Android applications,
and JVM experiments on the DaCapo 9.12-Bach benchmark
suite~\cite{dacapo:paper}.

\subsection{Android}

The first experiment compares the results of static analysis enhanced
with HeapDL output vs. plain, unenhanced context-insensitive static
analysis. We test a diverse set of Android benchmarks, chosen to be
realistic applications: Chrome, Instagram, S Photo Editor, Pinterest,
Google Translate, and Android Terminal Emulator.

We use Android 7.1, since it has support for heap profiling using the
same HPROF format as OpenJDK. We recompiled Android from sources
and produced two artifacts: (a)~a JAR containing all bytecode
corresponding to the Android Java API and (b)~an accompanying Android
virtual device image that can be loaded to the Android emulator. The
reason that we generated a custom JAR for the Android API is that the
JAR files that come with the Android SDK are stubs, i.e. they only
contain entry points to the API methods, plus some minimal
bytecode. This ``full'' JAR was given to the Doop static analysis
as the platform JAR and permits the analysis of the app on top of Android.

To dynamically exercise the applications, we ran our benchmarks in the
Android emulator with the UI/Application Exerciser Monkey
tool,\footnote{\url{https://developer.android.com/studio/test/monkey.html}}
which generates random input events, to simulate actual use of each
app. We performed at least 1024 random events. For some applications,
we conducted two runs: one to supply log-in credentials manually, and
another (after application shutdown and restart) to run with
Monkey. The heap dump was taken at the exact point Monkey sent the
last event. The statistical variability in Android is very low and is
eclipsed by other factors such as the number of random events and user
input.

Compared to later experiments, there are a few features of the Android
experiment to remember: \begin{inparaenum}
\item the baseline is a static analysis with static support for
  Android features but with no dynamic information---later
  experiments will compare with the Tamiflex tool, which currently is JVM-only;
\item there is no heap enrichment on the Android platform.
\end{inparaenum}

\begin{figure}

  \includegraphics[scale=0.32]{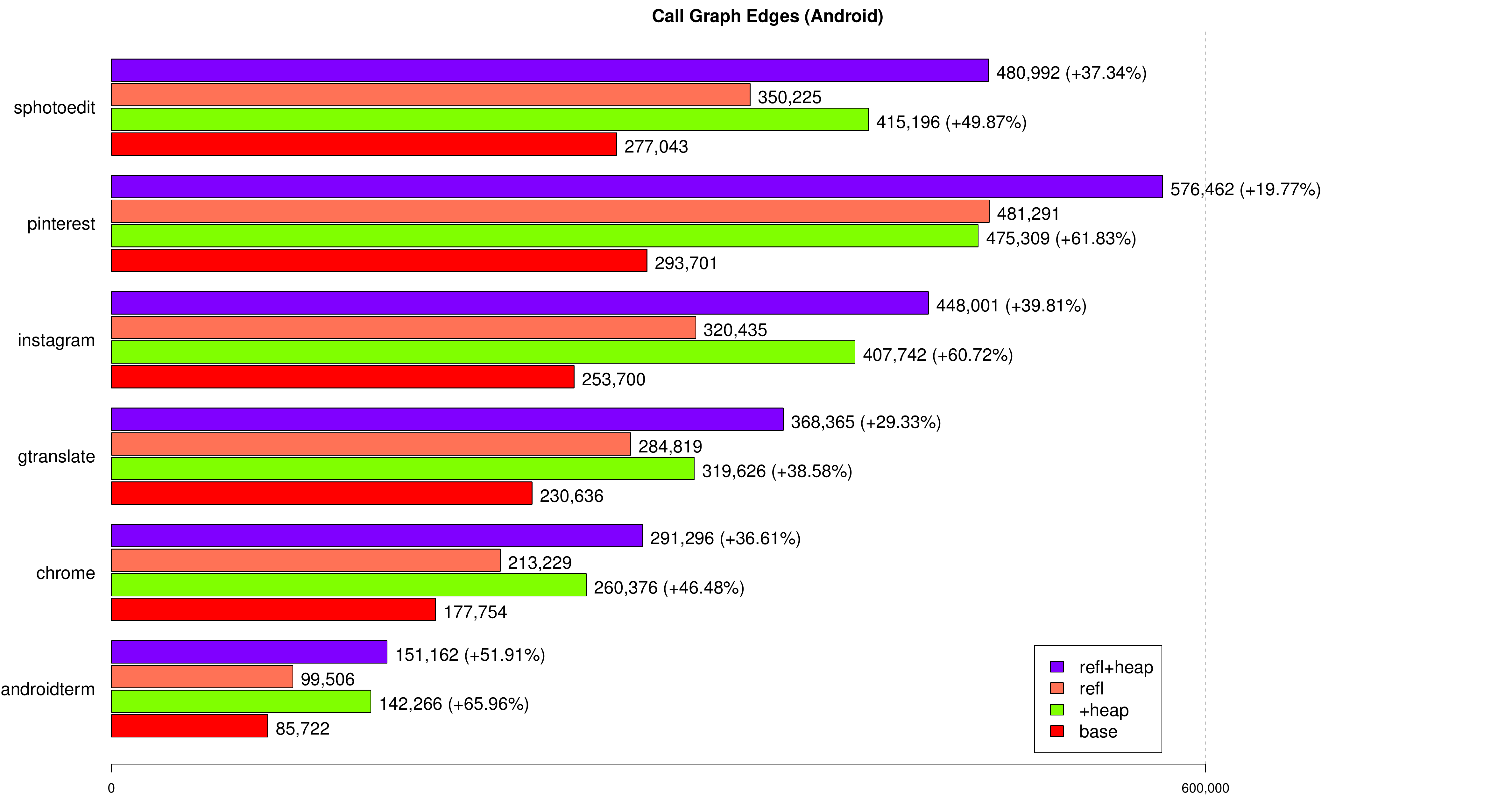}

  \caption{Android benchmarks: number of call-graph edges reported by the
    static analysis, with and without HeapDL assistance, with and
    without static reflection analysis.}
  \label{fig:android-callgraph}
\end{figure}

\begin{figure}
\includegraphics[scale=0.31]{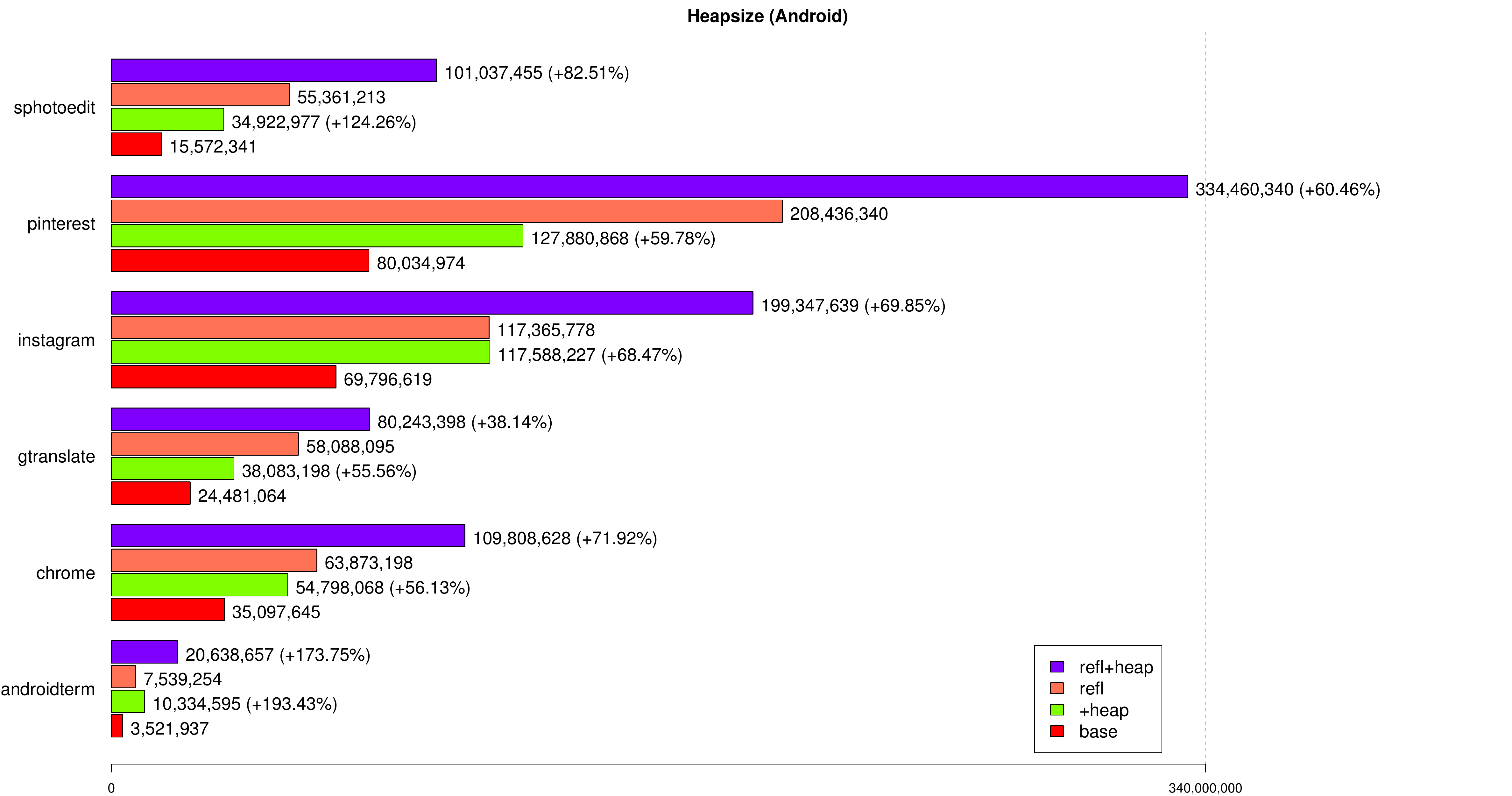}
  \caption{Android benchmarks: heap size reported by the
    static analysis, with and without HeapDL assistance, with and
    without static reflection analysis.}
  \label{fig:android-heap}
\end{figure}

The impact of HeapDL on the static analysis results (call-graph edges
and heap size) is shown in Figures~\ref{fig:android-callgraph}
and~\ref{fig:android-heap}. Every benchmark is run in four
configurations: no heap information or reflection support (base),
HeapDL information (+heap), static reflection support (refl), and both
reflection support and HeapDL information (refl+heap). Reflection
support is the ``classic-reflection'' mode of Doop, which analyzes
reflective features and reasons about strings used for reflection
purposes.

As can be seen, the increase in all metrics due to the use of heap
snapshots is drastic. Many tens of percent of extra call-graph edges
and an equal or larger increase in static heap size occur. The static
analysis reach expands significantly---the static analysis on its own
is not enough to discover this extra information, strongly suggesting
unsoundness. Even with static reflection analysis, the increase with
HeapDL input is large. All metrics support the position that HeapDL is
effective (\rqstyle{RQ.A1-3}). (The Android platform does not easily
permit dynamic instrumentation to see if the extra static analysis
results really capture valid dynamic behaviors---our later experiments
will address this.)

\begin{figure}
  \includegraphics[scale=0.33]{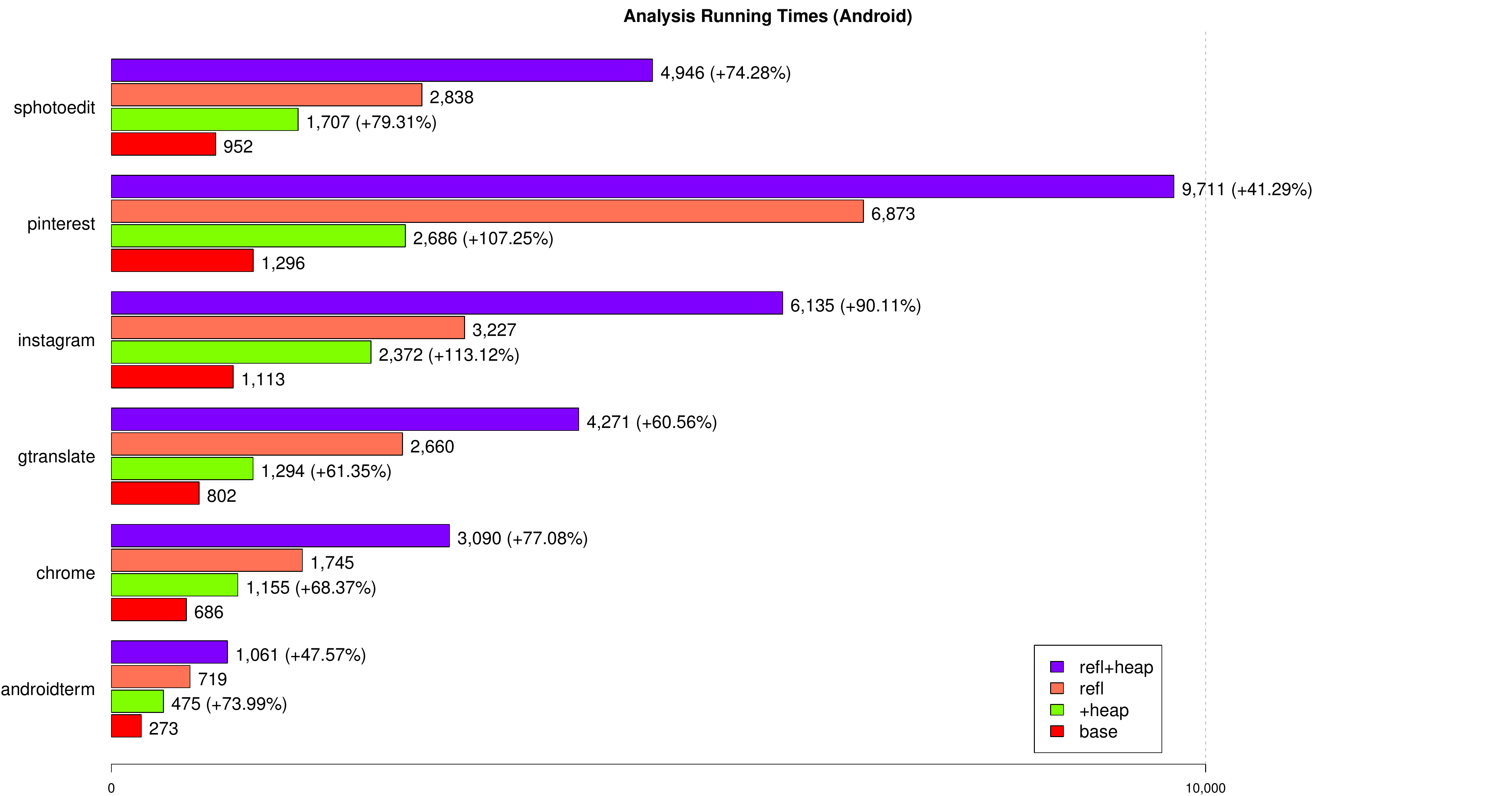}
  \caption{Analysis times for Android benchmarks.}
  \label{fig:android-times}
\end{figure}

Figure~\ref{fig:android-times} shows the running time of the static
analysis when enhanced with HeapDL input. The running time increase
(typically in the 60-80\% range) is commensurate with the increase in
overall analysis reach. Running times remain realistic, considering
how much more code is analyzed (as evident by the extra call-graph
edges) and the overall size of the programs involved---some of the
largest Android apps are included in our benchmark set. This suggests
that HeapDL is efficient and does not burden the static analysis
disproportionately (\rqstyle{RQ.B2}). 

Regarding the cost of the dynamic analysis (\rqstyle{RQ.B1}),
interestingly, we did not find a significant overhead when allocation
tracking is turned on in Android. Measuring application start-up,
runtime, shutdown times, and their sum, with and without allocation
tracking, we observed a significant variance (up to \% 10) between
runs of the same benchmark, but without any strong correlation with
allocation tracking being enabled. This could be due to the
interactive nature of most of the tested applications,\footnote{We had
  to set a delay of 2ms in the Monkey tool to avoid losing events.  If
  a mere 2ms interaction time is sufficient to hide profiling
  slowdown, it can be well argued that there is no perceptible
  slowdown in the first place.}  or to other overheads of the
system. We investigated this topic further by hand-crafting an Android
application that performs no I/O. Under this synthetic scenario we
observed a worst-case 48\% overhead of heap snapshots with allocation
tracking. It is telling that we had to resort to a synthetic benchmark
to obtain a measurable overhead.



\subsection{JVM Benchmarks: DaCapo}

Our second experiment examines the standard DaCapo 2009 Java benchmark
suite on the JVM.\footnote{We used version 1.8u131 of the Oracle JDK.}
We omitted a priori the Tomcat and Tradesoap benchmarks, to ease the engineering
requirements on our experimental setup. These benchmarks perform various kinds of
I/O, spawn webservers or other processes and are generally less ameanable to
profiling\footnote{\url{http://sourceforge.net/p/dacapobench/bugs/70/}}.

This experiment uses as baseline not a plain static analysis (as in
the previous section) but an analysis enhanced with dynamic
reflection information, produced by the state-of-the-art Tamiflex
tool. This is a key comparison for HeapDL. Our claim has been that
heap snapshots are an excellent way to compensate for the unsoundness
of static analysis, in a more complete way than merely recording
specific program actions (such as reflection calls).

The experiment is conducted with the heap enricher enabled, so that
the full dynamic call graph is registered (using the \sv{EdgeCtx}
objects), however the context information is ignored since the static
analysis is carried out context-insensitively. Also, the enricher for
capturing all dynamically loaded classes
(Section~\ref{sec:dynamic-code}) is not enabled, to avoid clouding the
results: that enricher does not affect the HeapDL performance much on
the DaCapo benchmarks and we want all analyses (static+Tamiflex
vs. +HeapDL) to run on the exact same bytecode.

The heap dump is taken on JVM exit. The application is instrumented by
the heap-enhancing agent to persist all relevant objects allocated by
the same application. There is negligible statistical variability in
the Dacapo benchmarks, since the harness is deterministic in the way
it exercises the application.

\begin{figure}
  \includegraphics[scale=0.31]{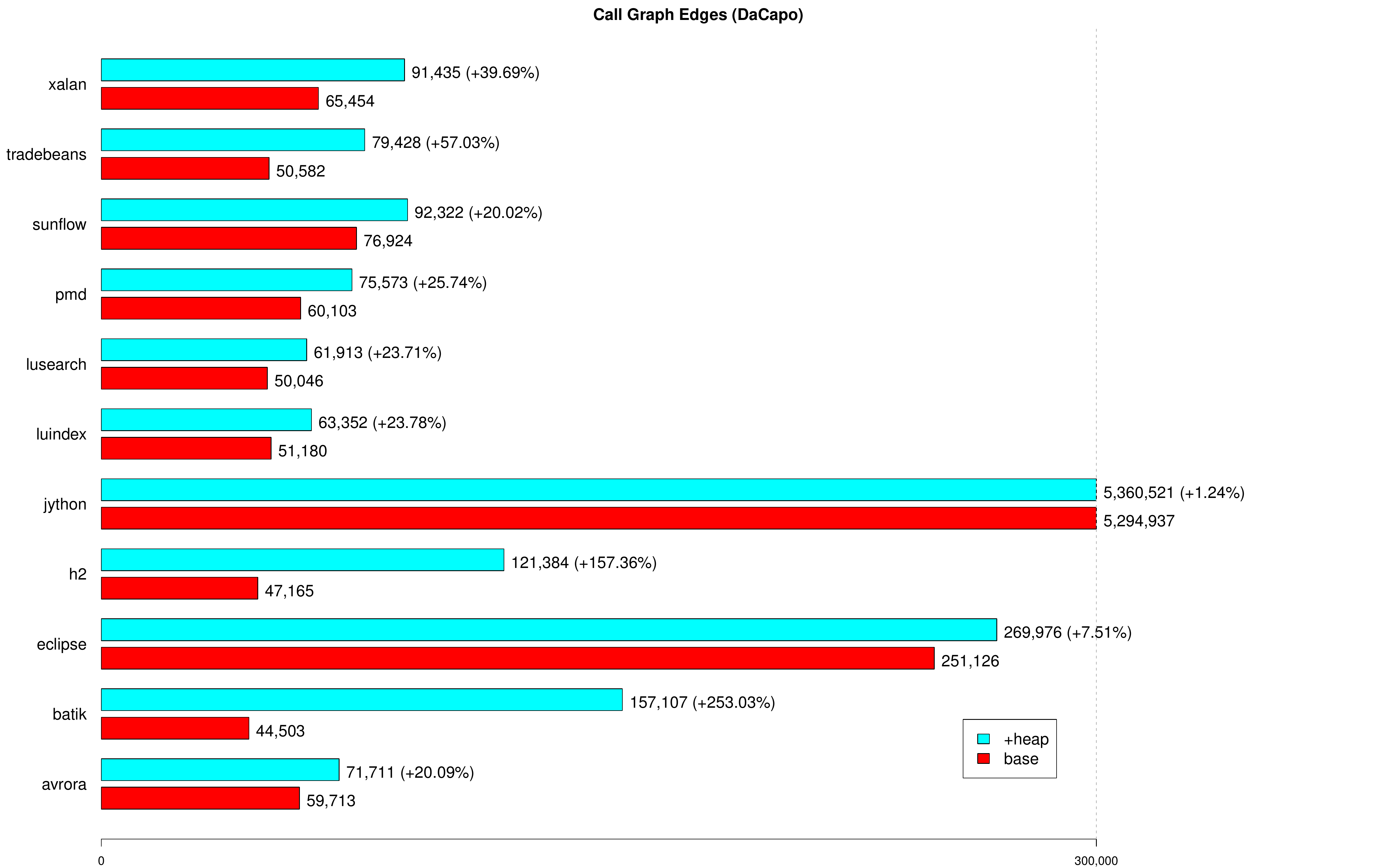}
  \caption[]{Call-graph size for DaCapo benchmarks. The figure is truncated for readability.}
  \label{fig:dacapo-callgraph}
\end{figure}

\begin{figure}
  \includegraphics[scale=0.31]{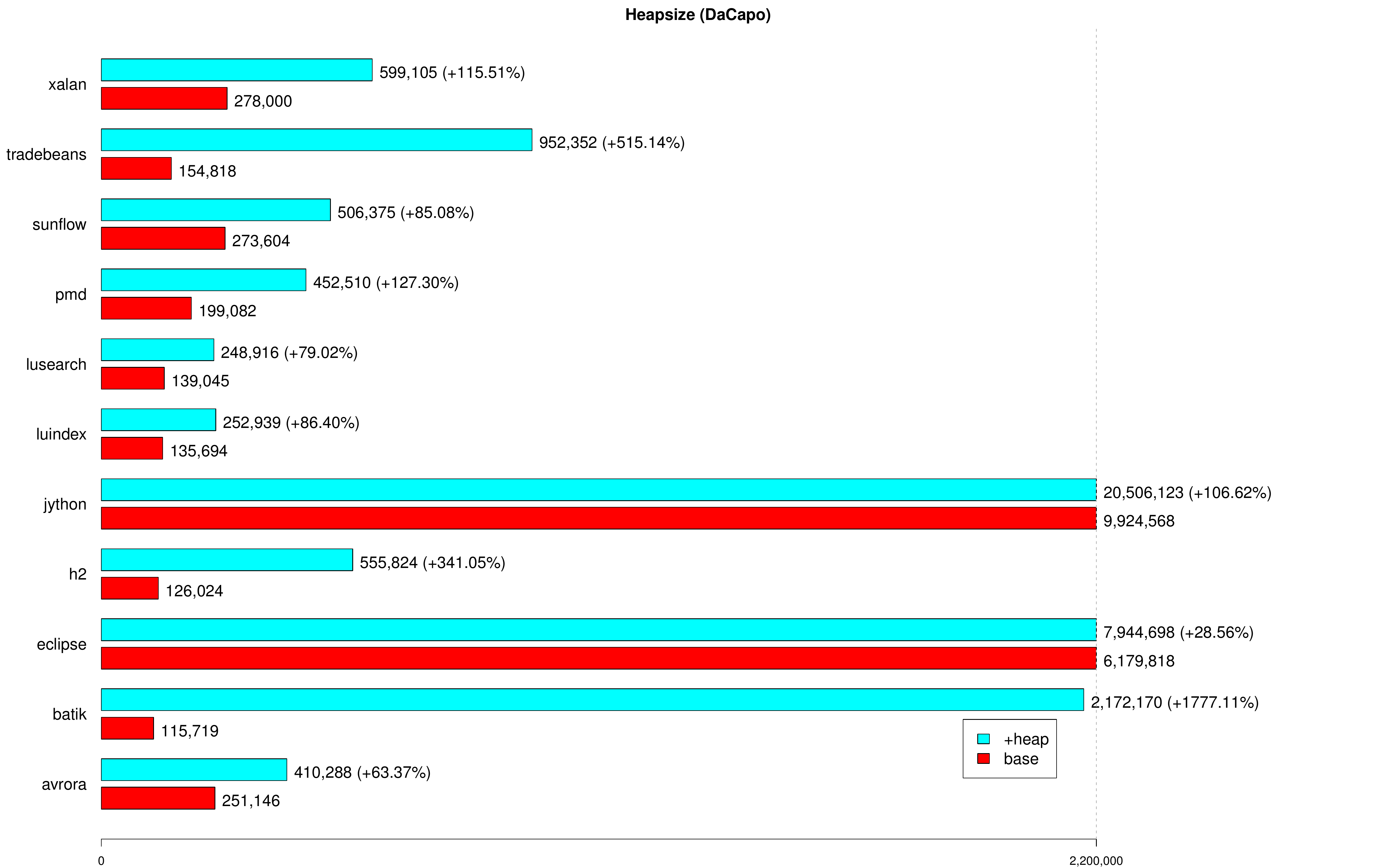}
  \caption{Heap size for DaCapo benchmarks. The figure is truncated for readability.}
  \label{fig:dacapo-heap}
\end{figure}

Figures~\ref{fig:dacapo-callgraph} and \ref{fig:dacapo-heap} show the
number of call-graph edges and heap size for the benchmarks. The two
bars are for the baseline static analysis (Doop with Tamiflex input)
with and without the HeapDL input. We used the ``default'' input size
of the DaCapo benchmarks for dynamic analysis.


As can be seen, the increase in analysis metrics is substantial,
typically at over-20\% more call-graph edges (median: 24\%), and even
higher for the size of the static heap (median increase: 86\%). The
call-graph edge increase is smaller than on the Android setting,
exactly as would be expected, since the Tamiflex input addresses some
of the unsoundness of the static analysis. Tamiflex is still missing
many call-graph edges, however.
(We note that the increase for the
batik benchmark is an outlier because Tamiflex misses a key call-graph
edge with the default input. Surprisingly, it observes it with the
``small'' input of the benchmarks. Thus, one should not consider
batik to be representative in terms of soundness, although it is still
informative in terms of other metrics.)
Therefore, on \rqstyle{RQ.A}, the experiment appears to strongly
confirm that HeapDL is effective and improves on the state of the art.

\begin{figure}
  \includegraphics[scale=0.31]{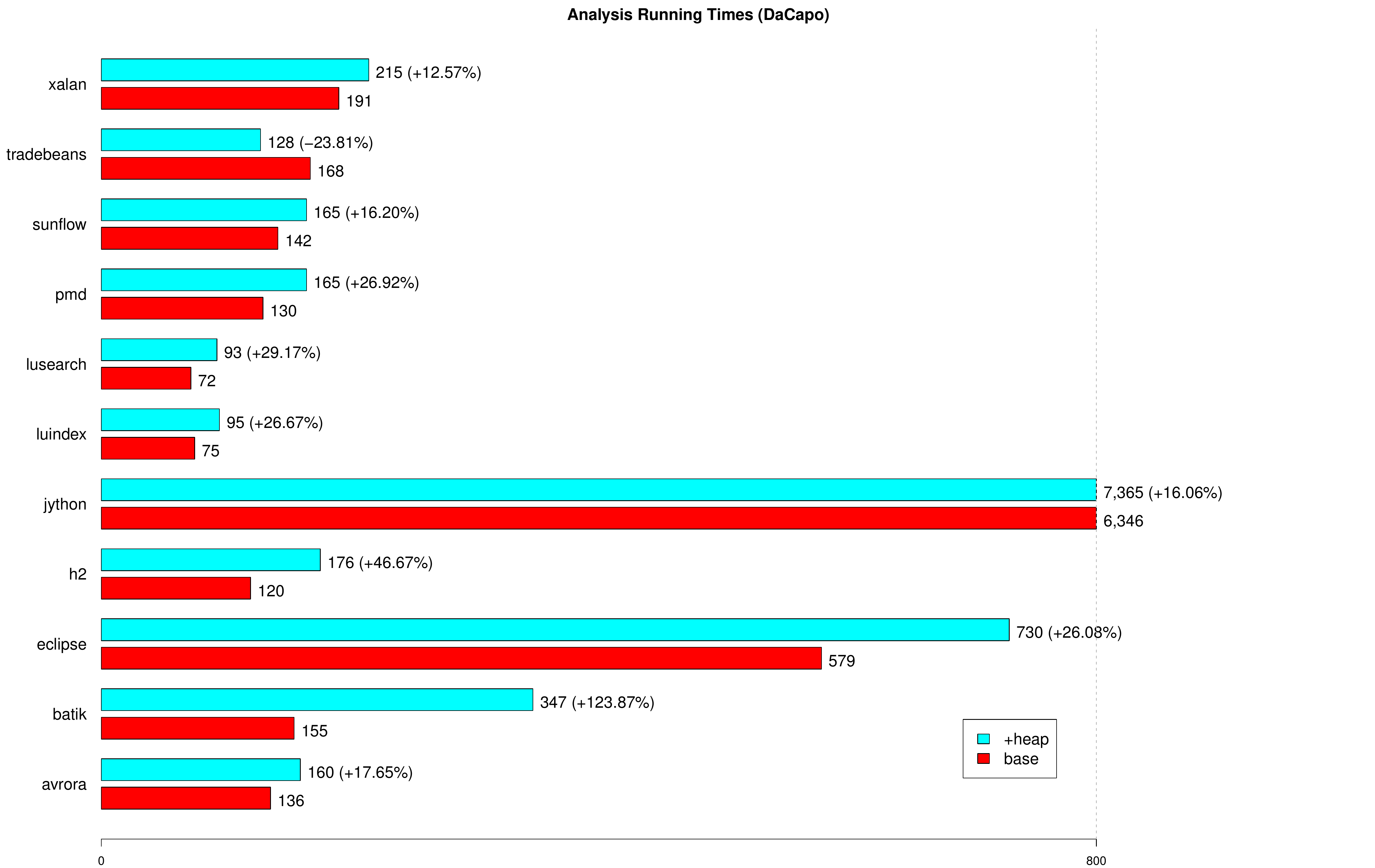}
  \caption{Analysis times for DaCapo benchmarks. The figure is truncated for readability.}
  \label{fig:dacapo-times}
\end{figure}

Figure~\ref{fig:dacapo-times} shows that the increase in analysis
reach comes with modest increases in static analysis cost
(\rqstyle{RQ.B2}). (We even see a surprising \emph{reduction}, for
tradebeans. We have not yet managed to explain this, but it is a
repeatable effect. We speculate that it is merely due to the analysis
reaching fixpoint a lot earlier due to the many initial dynamic
call-graph edges.)

\begin{figure}
  \includegraphics[scale=0.33]{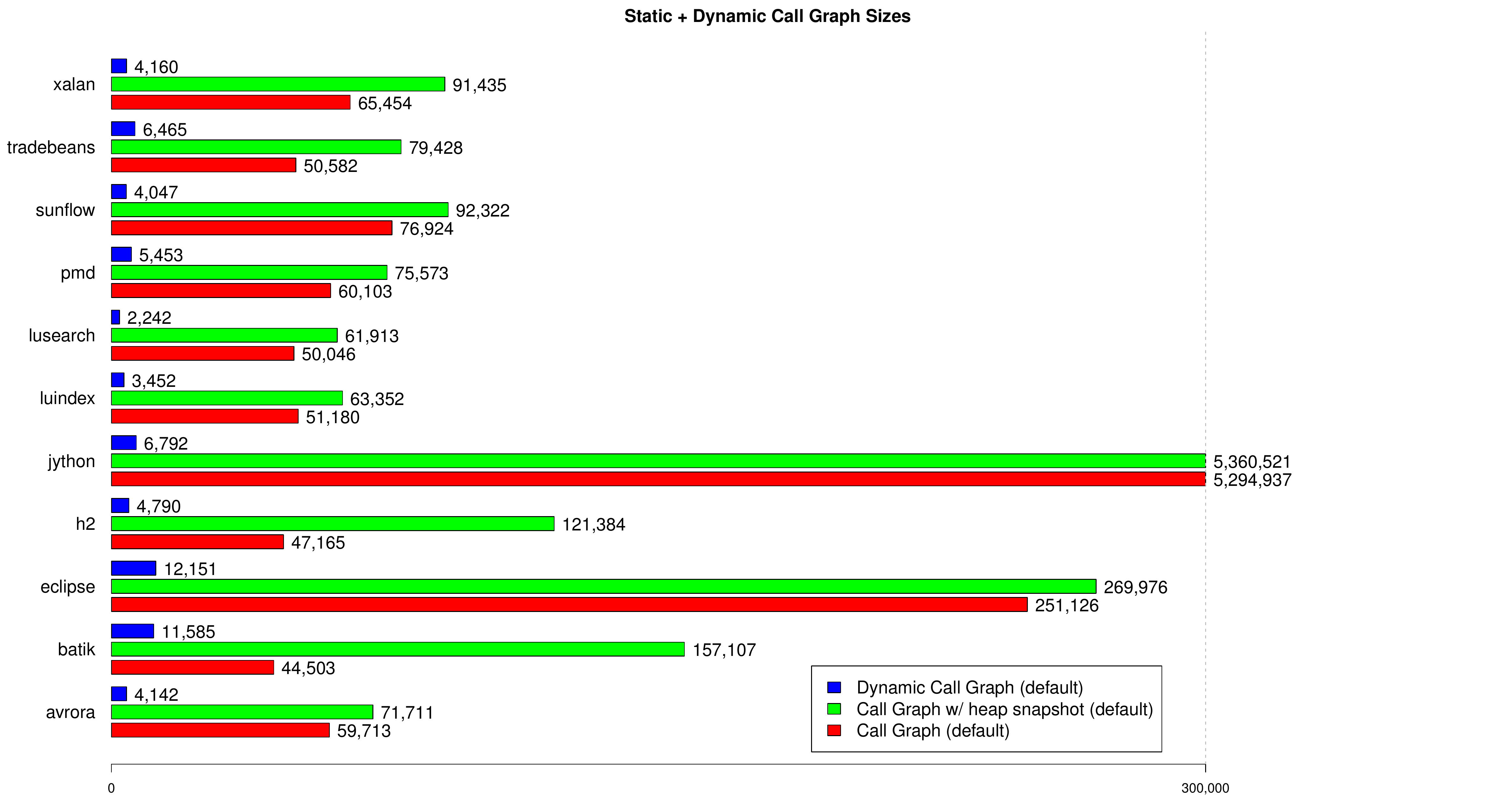}
  \vspace{-1em}
  \caption{Dynamic call-graph edges vs. increase in static call-graph edges. The figure is truncated for readability.}
  \label{fig:dacapo-dynamic}
\end{figure}

The extra time taken by static analysis when enhanced with HeapDL
inputs is a direct effect of enhancing the coverage of the analysis.
The information exposed by HeapDL makes the analysis infer more reachable
code, which in turn makes the analysis run longer. Indeed, the
information (e.g., all call-graph edges, not just new ones) that the static
analysis receives from HeapDL is a small proportion of the extra information that the
static analysis ends up inferring. This can be seen in
Figure~\ref{fig:dacapo-dynamic}, which plots the dynamic call-graph
edges produced by HeapDL against the (earlier-reported) call-graph
edges inferred by the static analysis with and without HeapDL. As can
be seen, the increase in static call-graph edges is typically 3-5x of
the dynamic call-graph edges that HeapDL provides.

On the other hand, the run-time cost (\rqstyle{RQ.B1}) is much higher
than on the Android platform. We have found that JVM-profiling with
allocation tracking on the DaCapo benchmarks incurs a 20-50x slowdown
(median: 39x). This slowdown is incurred for a standard configuration
of a reference profiler tool, so it is in line with overheads that
programmers observe during realistic profiling tasks.  Our optional
heap enriching agent compounds this cost with a further 1.1-10x
slowdown (median: 1.8x), for a total slowdown that can approach
two-to-three orders of magnitude! 

Thus, currently HeapDL pays a performance penalty on the dynamic
execution in order to yield inputs for enhanced static analysis.  We
expect that this cost is acceptable in the majority of cases. Dynamic
instrumentation often incurs high costs on high-performance platforms
and the overhead does not prohibit the actual execution of realistic
programs, when the stakes are as high as static (i.e., all-input)
analysis coverage.

We investigated tuning options in order to minimize the profiling
overhead. With a bounded depth of 6 for captured stack traces, the
analysis results are nearly identical to those reported in our full
experiments, yet the overhead of HPROF profiling drops to a median of
21x (instead of 39x). For future development, there are several
alternative profiler implementations that can potentially yield lower
overheads. These include the HPROF agent of the IBM
JDK~\cite{ibmhprof}, the YourKit profiler~\cite{yourkit}, the Java
Flight Recorder~\cite{jfr}, and the Java VisualVM
profiler~\cite{visualvm}. It is a strength
of the approach that profiling is done externally, by third-party
tools.

\subsection{Quantifying Coverage Increase}

A highly meaningful test for mechanisms that enhance the coverage of an
analysis is to measure their ability to anticipate \emph{unseen} behaviors.
We saw in Figure~\ref{fig:dacapo-callgraph} that HeapDL enhances a static
analysis to explore a lot more call-graph edges. But does this translate into
improved coverage of behaviors that truly arise? 

In order to measure the coverage increase (\rqstyle{RQ.C}) that
HeapDL enables, we compare the recall of the dynamic call-graph edges
for DaCapo executions under the ``default'' input size, \emph{when the
  static analysis has only seen the dynamic behavior of the ``small''
  input size}. That is, we first run the benchmarks with the ``small''
workload (for both the Tamiflex tool and HeapDL). This run serves to
produce inputs for the static analysis, which analyzes the program and
produces a static call graph. We then examine the recall of this
static call graph, against the dynamic call graph arising for an
execution with the ``default'' benchmark input.\footnote{We sought to
  perform the same experiment with the ``default'' vs. ``large''
  DaCapo inputs, but this is not available for the full set of
  benchmarks, and others fail for the supplied ``large'' input,
  without any instrumentation.}  The setup of the experiment is
otherwise identical as in the earlier DaCapo benchmark experiment,
i.e., we do not enable the enricher for dynamically-loaded code, so
that both Tamiflex and HeapDL operate on the same bytecode.



\begin{figure}
  \includegraphics[scale=0.34]{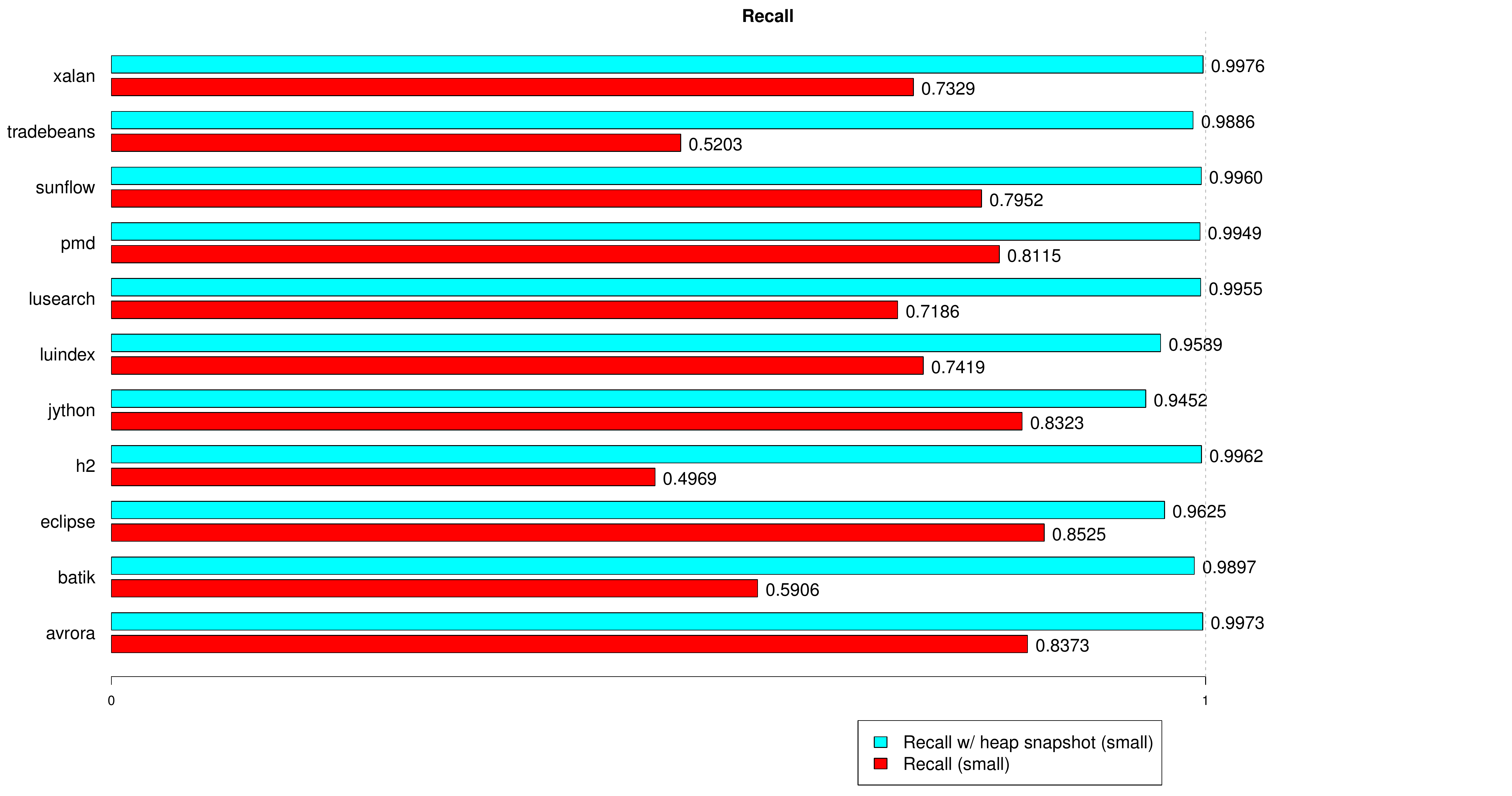}
  \vspace{-1em}
  \caption{Coverage/Recall for DaCapo benchmarks.}
  \label{fig:dacapo-soundness}
\end{figure}

Figure~\ref{fig:dacapo-soundness} shows the results of the recall comparison.
As can be seen, HeapDL results in a significant increase in call-graph
edge recall: the static analysis successfully infers almost all of the
dynamic call-graph edges in the ``default'' execution, which it has
never seen.  In contrast, the Tamiflex techniques alone are not enough
to achieve similar coverage: more than 20\% 
of the dynamic call-graph edges from differnt runs are missed. The median recall for the
baseline (static analysis + Tamiflex input) is 76.9\%, while it rises
to 99.5\% when HeapDL input is added. Thus, the experiment suggests
that the answer to \rqstyle{RQ.C} is affirmative: HeapDL increases
coverage of actual program behavior, compared to a state-of-the-art tool.

\subsection{Discussion}

Although we have not quantitatively classified the sources of
unsoundness in our experiments, we can share qualitative insights from
a manual inspection effort. Furthermore, we also show that the additional coverage
gained by using HeapDL directly translates to benefits in client analyses.

\subsubsection{Unsoundness in Static Analysis}
Many of the dynamic edges missed by the static analysis relate to
low-level code. This code, however, often translates into unsoundness
when analyzing the application itself. Our earlier figures
\ref{fig:dacapo-callgraph}, \ref{fig:dacapo-heap}, and
\ref{fig:dacapo-soundness} have 4 outliers among the DaCapo
benchmarks: batik, h2, tradebeans, and xalan. All four show
significant increase in the amount of application code found to be
reachable by the analysis. Without HeapDL input, the static analysis
often discovered less than 10\% of the code of these benchmarks to be
reachable. On the tradebeans benchmark, a substantial part of the
edges that are missed involve typical web server functionality, e.g.,
encryption, security, command-line parsing, etc.

An interesting observation is that the DaCapo benchmarks are older, so
they do not use the \invokedynamic{} instruction (used, e.g., in the
translation of lambda expressions). However, when analyzed in
conjunction with a Java 8 library, such instructions arise: the
benchmarks generate anonymous classes to be called via
\invokedynamic{} due to the automatic ``SAM conversion''. In new
JDKs, all single-abstract-method (SAM) types see invocations of their
methods transformed into \invokedynamic{}
calls~\cite{Goetz10,lambdas}. We have observed this arise in at least
three DaCapo benchmarks. Heap snapshots successfully compensate for
this semantic omission.

In the Android setting, an example of what we gain from heap data is
the discovery of many call graph edges to the graphical
subsystem. Apps in Android set up their user interface using a
``layout inflater'', which uses reflection via external XML files to
set up the GUI elements of the app. We found that such code is hostile
to classic static analysis (even with reflection analysis or Tamiflex
information).

\subsubsection{Benefits for Client Analyses}
HeapDL's effect of enhancing the call graph and reachable method
coverage of static analyzers directly translates to benefits for most
conceivable client analyses. For instance, for the Instagram
application on Android, a static taint analyzer~\cite{Grech2017infoflow}
flags 2.6x more suspicious information flows when enhanced with HeapDL
input.  This is a higher-than-proportional increase in possible
vulnerabilities flagged, relative to the metrics of
Figures~\ref{fig:android-callgraph} and \ref{fig:android-heap} on
Instagram (1.6x call-graph edges, 1.7x heap size increase).


This effect is hardly surprising. At a high level, a larger coverage
of reachable methods by a static analysis can easily translate into a
larger number of vulnerabilities detected, e.g., vulnerabilities may
lurk in the code that is not covered by the client analysis. At a more
detailed level, a larger coverage of call-graph edges yields
substantial increase in the behaviors covered by considering more
\emph{combinations} of events. For instance, in a taint analyzer, a
taint source, sink, or taint transfer method is often represented as a
method invocation. If the underlying analysis builds a larger, more
representative call graph, the numbers of all three elements (sources,
sinks, and taint transfer methods) will increase. Since an information
flow consists of combinations of such events, the increase will be
magnified.

\section{Related work}



The general pattern of adding dynamic analysis
information to address cases that are hard for static analysis has a
time-honored past, with approaches such as \emph{dynamic symbolic
  execution}~\cite{Godefroid05,Sen05} and environment models in model
checking. For example \citet{mercer2005model} present a model checking
approach that uses the GNU debugger to establish cycle-accurate
effects of the compiled program elements under different backends and
processors.  What distinguishes our approach is the use of heap
snapshots with allocation tracking, as well as the emphasis on the
information (e.g., dynamically-loaded code, object-sensitive contexts)
that is particularly valuable for a whole-program static analysis.

\citet{Li17} show how to combine dynamic symbolic execution with
subtype polymorphism in Java to resolve the targets of method
invocations. Their approach improves soundness relative to plain
dynamic symbolic execution but does not address the soundness issues
relative to native code, heterogeneous applications, or \invokedynamic{}.

\citet{ecoop/HirzelDH04} show one of the first works that consider
runtime monitoring so as to obtain information for state-of-the-art
Java program analysis techniques. Concretely, they extend Andersen's
pointer analysis algorithm to an online setting, which enables it to
handle dynamic class loading, reflection and native code (through the
disciplined JNI interface). The system observes such events and
re-runs the analysis with these observations taken into account.  The
approach is conceptually closely related to ours, since it targets the
same kinds of analyses (whole-program points-to and
call-graph). However, the Hirzel et al. approach is quite different in
its characteristics: it requires full control of the runtime
environment; it applies only to analyses that are inexpensive enough
to re-run regularly; it does not separate the dynamic information from
the static analysis; it relies on capturing events and not effects
(e.g., it will not intercept low-level heap updates through unsafe
APIs).

Dynamic class loading with reflection is a hard problem for static
analysis: \citet{Landman17} note that there are still soundness problems
in the handling of dynamic proxies or reflection, even
with state-of-the-art techniques. Our approach partly addresses the
shortcomings of these past techniques by using run-time heap
information to detect behaviors that would otherwise be missed. There are
also techniques that attempt to statically analyze either dynamic
class loading or reflection. For example, \citet{Yoshiura14} do static
data race detection in the presence of dynamic class loading (an analysis that
is not fully automatic as it requires manual handling of loops) and
\citet{Li2015} improve the static handling of reflection, but do not
fully handle native methods, \invokedynamic{}, or heterogeneous code.

Inspecting a program's state prior to static analysis is a strategy sometimes
employed by hybrid static analysers for dynamically typed languages. For
instance, RPython~\cite{ancona2007rpython} lazily inspects most of the global state, and
combines dynamic and static type information to perform a flow-sensitive type
inference, before generating optimized code. A similar implementation strategy
is used in preemptive type checking~\cite{grech2013preemptive}. On the other
hand, the use of profiling on hard to analyse features is used in
PRuby~\cite{furr2009profile}, particularly on Ruby's \texttt{eval}
and other unsafe functions to infer possible side effects of these.

The Tamiflex tool~\cite{icse/BoddenSSOM11} (extensively discussed earlier) employs a \emph{`Play-out'}
agent to log runtime reflective calls and classes loaded via custom
class loaders or on-the-fly code generation.  A secondary tool
component called the \emph{'Booster'} then enriches the respective
classes, at the point where the reflective calls are made, by inserting regular method
calls that materialize the reflective calls, thus making them
detectable by standard static analyzers.  The Booster component also
instruments runtime checks that warn the user when the analyzed program
executes reflective calls that were not executed in the previous runs
(and may, thus, be a source of unsoundness).  The Tamiflex toolchain
also provides support for inserting offline-transformed classes into a
running program via a \emph{`Play-in'} agent. All such functionality would
be interesting to incorporate in the HeapDL tool in the future.

Our approach is certainly not the first to recognize the high value of
heap snapshots. Specifically in Java, there have been several research
uses of HPROF data---e.g., most directly for different kinds of heap
visualization~\cite{Aftandilian10,Reiss09}. Other dynamic analyses use
heap profiling data to check aliasing properties~\cite{Potanin04},
analyze synchronization performance~\cite{Hofer15}, generate software
birthmarks~\cite{Chan11}, or diagnose memory leaks~\cite{Maxwell10}.


The problem of static analysis unsoundness is particularly acute for
Android frameworks, since they make heavy use of
reflection. Droidel~\cite{Blackshear15} simulates some uses of
reflection in Android and replaces reflective behavior with static
calls to generated code (stubs); these can then be processed using
off-the-shelf analysis tools for Android apps (e.g., Soot and
WALA~\cite{www:wala}) that would otherwise miss these call-graph
edges.  Their approach is however \emph{not} fully automated and
requires manual code instrumentation for explicating reflection. In
HybriDroid \cite{Lee:2016:HSA:2970276.2970368}, dynamic analysis is
not employed to model the effects of foreign code. Instead, the
interaction between Android and JavaScript code is handled
explicitly. This means that standard WebView browser component events
are explicitly modeled and JavaScript code running in this is analyzed
together with the application.

\citet{Zhauniarovich15} observe that an \emph{``extensive amount of
  Android apps relies on dynamic code update features''} and offer a
combination of static and dynamic approaches for security analyses on
Android. In particular, the dynamic part of their approach is using a
modified Android operating system that runs the application to be
analyzed. In comparison, our work differs in that it is
non-intrusive. We do not need a modified version of Android to take
heap snapshots (although we built a platform image from scratch to
have a tightly controlled environment for benchmarks in this
paper).

\section{Conclusions} 

We presented an approach to enhancing the coverage of a static
analysis by employing dynamic information. Although the general
pattern for such an enhancement is well-established, our techniques
are interesting in their specifics. We use modern heap-snapshot and
allocation-tracking technology provided by profiling APIs in
mainstream platforms; we export dynamic information in a format
suitable for whole-program static analysis, such as call-graph
analysis and pointer analysis of a global heap; we enrich the heap for
various purposes, including maintaining context information that the
static analysis expects. Our approach is embodied in the HeapDL tool,
which we show can achieve significant increases in analysis 
coverage, compared to the closest baselines.

We believe that heap snapshots are the right tool for addressing
unsoundness shortcomings of static analyses. Heap snapshots offer the enormous
advantage of being non-intrusive: there is no need to instrument code
(except for purposes of getting optional extra information, as in heap
enrichment) or, generally, to watch for specific program
actions. Instead, a wealth of information on a program's behavior is
readily available by observing the effects of the program on the heap
and snapshot call-graphs. Such program effects can capture semantic
elements of even the hardest-to-analyze code: native actions,
dynamically generated code, and all sorts of unsupported
functionality.

\begin{acks}
  We gratefully acknowledge funding by the European Research
Council, grant 307334 (SPADE). In addition, the research work disclosed is
partially funded by the REACH HIGH Scholars Program -- Post-Doctoral
Grants. The grant is part-financed by the European Union, Operational Program
II, Cohesion Policy 2014-2020 (Investing in human capital to create more
opportunities and promote the wellbeing of society - European Social Fund).
\end{acks}


\bibliography{bibliography,bib/ptranalysis,bib/proceedings,bib/tools}

\end{document}